\documentclass{article}

\usepackage{arxiv}

\usepackage{graphicx}
\usepackage{xcolor}
\usepackage{amsmath}
\usepackage{amsfonts}
\usepackage{amsthm}
\usepackage{bm}
\usepackage{array}
\usepackage{booktabs}
\usepackage[separate-uncertainty=true,table-align-uncertainty=true]{siunitx}
\usepackage[caption=false]{subfig}
\usepackage{placeins}
\usepackage{multirow}
\usepackage{hyperref}
\DeclareMathSymbol{\IKap}{\mathalpha}{letters}{"14}
\newcommand{\GKap}{\boldsymbol{\IKap}}

\newtheorem*{remark}{Remark}

\title{Uncertainty Quantification in Calibration and Simulation of Thermo-Chemical Curing of Epoxy Resins}

\date{}

\author{
    \href{https://orcid.org/0000-0002-4999-4558}{\includegraphics[scale=0.06]{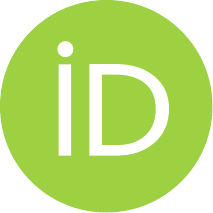}\hspace{1mm}Jendrik-Alexander Tröger} \\
	Institute of Applied Mechanics\\
	Clausthal University of Technology\\
	Adolph-Roemer-Stra{\ss}e 2a \\
	38678 Clausthal-Zellerfeld, Germany \\
	\texttt{jendrik-alexander.troeger@tu-clausthal.de} \\
    \And
    \href{https://orcid.org/0009-0000-7809-9889}{\includegraphics[scale=0.06]{orcid.pdf}\hspace{1mm}Christina Steinweller} \\
	Institute of Applied Mechanics\\
	Clausthal University of Technology\\
	Adolph-Roemer-Stra{\ss}e 2a \\
	38678 Clausthal-Zellerfeld, Germany \\
	\texttt{christina.steinweller@tu-clausthal.de} \\
	\And
	\href{https://orcid.org/0000-0003-1849-0784}{\includegraphics[scale=0.06]{orcid.pdf}\hspace{1mm}Stefan Hartmann} \\
	Institute of Applied Mechanics\\
	Clausthal University of Technology\\
	Adolph-Roemer-Stra{\ss}e 2a \\
	38678 Clausthal-Zellerfeld, Germany \\
	\texttt{stefan.hartmann@tu-clausthal.de} \\
}

\renewcommand{\headeright}{}
\renewcommand{\undertitle}{}
\renewcommand{\shorttitle}{Uncertainty Quantification of Thermo-Chemical Curing Processes}

\hypersetup{
pdftitle={Uncertainty Quantification in Calibration and Simulation of Thermo-Chemical Curing of Epoxy Resins},
pdfsubject={},
pdfauthor={Jendrik-Alexander Tröger, Christina Steinweller, Stefan Hartmann},
pdfkeywords={first-order second-moment method, Monte Carlo method, model calibration, curing, shrinkage, epoxy resins},
urlcolor=black,
citecolor=blue
}

\begin{document}
\maketitle

\begin{abstract}
    Curing of epoxy resins poses a particular challenge in terms of modeling, experimental investigation, and numerical implementation, as it is a thermo-chemo-mechanical process.
    Several constitutive relations are required to model these processes, yielding numerous material parameters. The calibration of the constitutive relations must be performed using multiple steps, wherein uncertainties unavoidably propagate. In this study, we investigate the propagation of uncertainties during both the multi-step calibration procedure and the numerical simulation of curing processes with the identified parameters. For both, we employ the first-order second-moment method, which is carefully evaluated through coverage tests and by comparing it to the Monte Carlo method as a reference. It is demonstrated that the first-order second-moment method efficiently yields reasonable results, although providing only a first-order approximation of the highly nonlinear stochastic model response.
\end{abstract}


\section{Introduction and Motivation}
\label{sec:introduction}

\begin{figure}
    \centering
    \includegraphics[width=0.25\linewidth]{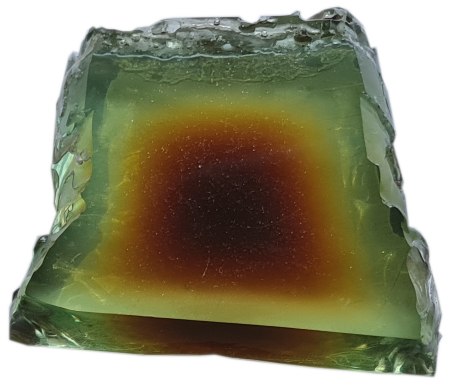}
    \caption{Local degradation due to overheating during exothermal chemical reaction}
    \label{fig:epoxydegraded}
\end{figure}

In the last two decades, composite materials have garnered significant interest in both research and industry, primarily due to their suitability for lightweight applications in the aviation industry, see \cite{ramonsguazzomoreira2018}. Within the present work, we focus on composite materials composed of a polymer matrix and reinforcing fibers, as opposed to metallic multi-material composites. Naturally, the polymer matrix is an essential component, which is typically a thermosetting polymer. To be more specific, epoxy resins are particularly well-suited due to their high mechanical strength and temperature resistance, arising from building extensively cross-linked three-dimensional networks \cite{shundoyamamototanaka2022}. During the manufacturing process, for instance, resin transfer molding, pultrusion, or filament winding, controlling the curing behavior of the epoxy resin is essential to achieve the desired component properties. The curing process itself, which is the solidification of the initially liquid epoxy resin, is based on an exothermal reaction. This exothermal nature of the chemical reaction causes inhomogeneous temperature distributions in thick-walled parts due to the low thermal conductivity of epoxy resins. Consequently, the curing process must be controlled to prevent local temperature hot spots, inhomogeneous curing of matrix material, or local degradation, visualized in Fig.~\ref{fig:epoxydegraded}. Another important aspect is the shrinkage behavior of the epoxy resin during curing because shrinkage in conjunction with inhomogeneous temperature distributions can lead to residual stresses or delamination, as shown in \cite{rabearisonjochumgrandidier2011}. Accordingly, we numerically investigate the thermo-chemical curing behavior of epoxy resins in this contribution. The constitutive relations were initially developed in fundamental work by \cite{leistnerhartmannablizziegmann2020} and subsequently extended in \cite{leistnerdiss2022}. However, uncertainty propagation was neglected in the aforementioned studies. We will explore this aspect during the multi-step calibration procedure and subsequent numerical simulations. 

The calibration of constitutive models is equally possible with deterministic and stochastic procedures. For a comprehensive overview focusing on model calibration in computational solid mechanics, we refer to \cite{roemerhartmanntroegerantonwesselsflascheldelorenzis2025}. Calibrating the thermo-chemical model for epoxy resins requires a multi-step procedure exploiting different sets of experimental data, each sufficiently addressing the specific model response. Additionally, information about measurement noise is not available for all calibration steps, as only a few data points were measured without repetitions. Thus, in this work, we employ the ordinary nonlinear least-squares method for model calibration. At first glance, the deterministic nature of the least-squares method may seem to prohibit uncertainty quantification and propagation during multi-step calibration. However, material parameter uncertainties can be estimated using the asymptotic properties of the nonlinear least-squares estimator, as derived by \cite{jennrich1969}, and applied in the context of model calibration by \cite{troegerroemerhartmann2024}. Since some calibration steps rely on previously identified parameters within the multi-step scheme, we utilize the first-order second-moment method (FOSM) to propagate these parameter uncertainties. 

Once the material parameters and their associated uncertainties have been determined using the multi-step calibration scheme, an important question arises: How do these parameter uncertainties influence the model response in numerical simulations of the curing and shrinkage behavior of epoxy resins? Although the model itself is deterministic, the presence of uncertain material parameters as inputs inevitably renders the model response uncertain as well. A large variety of methods is available to compute the model response uncertainties, see \cite{stefanou2009} for a review. Without intention of being an exhaustive list, available methods include:
\begin{itemize}
	\item Sampling-based methods \cite{caflisch1998,liu2004}
	\item Stochastic collocation methods \cite{xiuhesthaven2005,dannertetal2022}
	\item Stochastic Galerkin methods/polynomial chaos expansions \cite{ghanemspanos1991,xiukarniadakis2002}
	\item Perturbation methods \cite{kaminski2013,troegerkulozikhartmann2025}
	\item Time-separated stochastic mechanics \cite{junkernagel2020,geisleretal2025}
\end{itemize}
Sampling-based methods, such as the Monte Carlo method and related techniques, are widely recognized for their robustness, as they only rely on repeated model evaluations with varying input parameters. Despite this robustness, these methods are often less attractive due to their slow convergence rate. Although the convergence rate of the Monte Carlo method is independent of the problem's dimensionality, the number of samples required to accurately resolve typically nonlinear input-output relationships can become very large. As a result, the computational cost may become prohibitive, even if individual model evaluations are relatively fast or parallelization is exploited.

In contrast, stochastic collocation methods are an appealing approach due to their non-intrusive nature and rapid convergence. However, as noted in \cite{dannertetal2022}, stochastic collocation methods can incur high computational costs and exhibit instability when applied to nonlinear constitutive relations. Consequently, stochastic collocation methods are unsuitable for the present study, which involves such nonlinearities.

Stochastic Galerkin methods have attracted considerable research interest since their introduction by \cite{ghanemspanos1991}. Within this framework, non-intrusive polynomial chaos expansions have become popular for handling strong parameter variations and achieving high convergence rates. Nevertheless, these expansions still require multiple model evaluations to determine the coefficients, which can be challenging for high-dimensional input spaces. In contrast, perturbation methods based on Taylor expansions of the model response are computationally efficient, but are generally limited to simpler algebraic constitutive relations and small input parameter uncertainties \cite{kaminski2013,kaminski2022}. The FOSM used here is one such perturbation method. It is well established in design optimization, see exemplarily \cite{kruegeretal2023}, but has only recently been applied to model calibration \cite{dileephartmannetal2022,delizetal2025} and extended to ordinary differential equations \cite{troegerkulozikhartmann2025}.

A relatively recent approach for capturing the effects of uncertain material parameters on model outputs is time-separated stochastic mechanics, originally introduced by \cite{junkernagel2020}. The method has since been extended and applied to various constitutive models, see \cite{geisleretal2025} and the literature cited therein. In brief, time-separated stochastic mechanics involves reformulating the constitutive model to derive evolution equations that track the effects of material parameter fluctuations, analogous to the well-established evolution equations for internal variables in constitutive modeling. All sources of uncertainty are then incorporated in a post-processing step. Time-separated stochastic mechanics shares certain similarities with perturbation methods, as it also employs Taylor expansions. However, the process of tracking stochastic effects through time integration of differential equations must be handled carefully. The time-separated stochastic mechanics approach facilitates this process, albeit at the cost of deriving additional evolution equations.

In this contribution, we employ the FOSM as a throughout approach for efficient and sampling-free uncertainty quantification in multi-step model calibration of nonlinear constitutive relations and even during uncertainty quantification of numerical simulation results. The FOSM is evaluated for both the inverse problem and the forward simulation and assessed considering the reduced accuracy due to the first-order approximation by comparison with the Monte Carlo method.

We begin by briefly introducing the constitutive relations describing the thermo-chemical curing process of epoxy resins in Sec.~\ref{sec:constRelations}, along with a recap of the underlying experimental investigations. Next, in Sec.~\ref{sec:modelCalib}, we explain the applied multi-step calibration approach. The uncertainty quantification in Sec.~\ref{sec:uq_calib} addresses the propagation of uncertainties throughout the calibration procedure. Finally, we study the impact of material parameter uncertainties on numerical simulations of the curing process of an epoxy resin in Sec.~\ref{sec:uq_model}. The effectiveness of the FOSM for this task is demonstrated by comparison with the Monte Carlo method as a reference.

\section{Constitutive Relations for Thermo-Chemical Curing}
\label{sec:constRelations}
\begin{figure}[ht]
    \centering
    \includegraphics[width=0.65\linewidth]{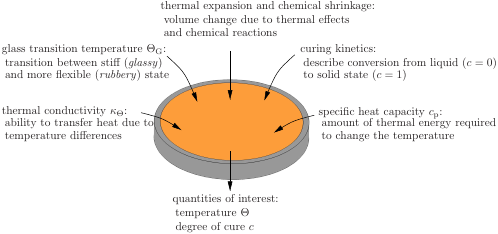}
    \caption{Schematic illustration of the physical quantities involved in the thermo-chemical analysis of curing processes}
    \label{fig:physQuantities_curing}
\end{figure}

The constitutive modeling of polymer curing has received significant research interest in the continuum mechanics community, as exemplified by \cite{lionhoefer2007,hossainsteinmann2015}.
The constitutive relations used in this paper for describing the curing and shrinkage behavior of epoxy resins during the exothermal curing reaction were developed in \cite{leistnerhartmannablizziegmann2020}. The shrinkage relation was slightly modified and extended to consider the glass transition temperature \cite{leistnerdiss2022}. Notably, the curing epoxy resin is modeled assuming a homogeneous medium. In more detail, the constitutive relations comprise model equations for: 1) glass transition temperature, 2) curing kinetics, 3) thermal expansion and chemical shrinkage, 4) specific heat capacity, and 5) thermal conductivity. The majority of the relations depend on the temperature $\mathrm{\Theta}$ and degree of cure $c$, which is a dimensionless number between $0$ (uncured material) and $1$ (fully cured material), indicating the extent of polymerization. The physical quantities relevant for modeling the thermo-chemical curing process and their meanings are illustrated in Fig.~\ref{fig:physQuantities_curing}.

\paragraph{Glass transition temperature.} The glass transition temperature $\mathrm{\Theta}_\mathrm{G}$ evolves when the curing proceeds and is itself influencing curing kinetics, specific heat capacity, and chemical shrinkage behavior. An overview of various models for the glass transition temperature is provided by \cite{urbaniak2011}. Here, we choose the well-known DiBenedetto equation,
\begin{equation}
	\label{eq:thetaG}
	\mathrm{\Theta}_\mathrm{G}(c) = \frac{r_\mathrm{f}\,c}{1-(1-r_\mathrm{f})c}\left(\mathrm{\Theta}_\mathrm{G1} - \mathrm{\Theta}_\mathrm{G0}\right) + \mathrm{\Theta}_\mathrm{G0},
\end{equation}
where $r_\mathrm{f}$ is a curvature parameter and $\mathrm{\Theta}_\mathrm{G0}$ and $\mathrm{\Theta}_\mathrm{G1}$ represent the glass transition temperature of the uncured and fully cured epoxy resin. The material parameters are identified using heat flux data from differential scanning calorimetry (DSC). The DSC data yield experimental data of the glass transition temperature with respect to the degree of cure using the definition $c = h/h_\mathrm{u}$ with $h$ denoting the reaction enthalpy and $h_\mathrm{u}$ being the ultimate reaction enthalpy. The glass transition temperature is characterized by a pronounced change in the specific heat capacity profile.

\paragraph{Curing kinetics.} The Kamal-Sourour model \cite{kamal74} is the most widespread model for curing evolution. The model includes an autocatalytic reaction term with an $n$th-order curing reaction, in which the temperature-dependent curing kinetics are described by the Arrhenius equation. The original model was essentially extended by \cite{coleetal1991,fournieretal1996} to consider that the initially chemically-driven curing reaction changes to a diffusion-driven reaction when the evolving glass transition temperature $\mathrm{\Theta}_\mathrm{G}$ reaches the curing temperature. After extensively investigating parameter correlation and uniqueness, \cite{leistnerhartmannablizziegmann2020} proposed a simplified version of the Kamal-Sourour model,
\begin{equation}
	\label{eq:curingKin}
	\dot{c} = f_\mathrm{c}(\mathrm{\Theta},c)f_\mathrm{d}(\mathrm{\Theta},c)
\end{equation}
with a chemically-driven term
\begin{equation}
	\label{eq:fc}
	f_\mathrm{c}(\mathrm{\Theta},c) = A\exp\left(-\frac{E}{R\mathrm{\Theta}}\right)\left(g + (1-g)c-c^2\right)^n
\end{equation}
and a diffusion-driven term
\begin{equation}
	\label{eq:fd}
	f_\mathrm{d}(\mathrm{\Theta},c) = \frac12 \left(1-\tanh\left(\frac{\mathrm{\Theta}_\mathrm{G}(c)-\mathrm{\Theta}}{b_\mathrm{d}}\right)\right).
\end{equation}
In Eq.~\eqref{eq:fc}, $R = \SI{8.314}{\J\per\K\per\mol}$ is the universal gas constant. The material parameters of the chemically-driven part of the curing kinetics are the pre-exponential factor $A$, activation energy $E$, a dimensionless factor $g$, and the dimensionless order exponent $n$. The diffusion-driven term \eqref{eq:fd} contains only the diffusion parameter $b_\mathrm{d}$ as a material parameter, see \cite{leistnerhartmannablizziegmann2020} for details. The material parameters are identified in two steps, distinguishing the parameters of the chemically-driven part \eqref{eq:fc} and the diffusion-driven part \eqref{eq:fd}. Experimental data are taken from isothermal DSC measurements, whereas density data can be employed as well, leading to comparable results \cite{leistnerdiss2022}.

\paragraph{Thermal expansion and chemical shrinkage.} The constitutive relation for the thermo-chemical deformation behavior of the curing epoxy resin comprises several physical effects. The thermal expansion of the uncured material is simply a linear function, $J_\mathrm{\Theta}(\mathrm{\Theta}) = 1 + \alpha_\mathrm{\Theta}\vartheta$. Here, $\alpha_\mathrm{\Theta}$ is the thermal expansion coefficient and $\vartheta = \mathrm{\Theta} - \mathrm{\Theta}_\mathrm{R}$ the temperature difference with reference temperature $\mathrm{\Theta}_\mathrm{R} = \SI[round-precision=0]{20}{\celsius}$. Note that the deformations are formulated as $J = \rho_\mathrm{R}/\rho$, representing the ratio between reference density $\rho_\mathrm{R}$ and actual density $\rho$. The chemical shrinkage during the curing process is modeled in \cite{leistnerhartmannablizziegmann2020} using additional material parameters, namely, the chemical shrinkage coefficient $\alpha_\mathrm{c}$ and the coupling parameter $\alpha_{\mathrm{\Theta}\mathrm{c}}$. This leads to the relation for thermo-chemical deformations
\begin{equation}
J_{\mathrm{\Theta}\mathrm{c}}(\mathrm{\Theta},c) = 1 + \alpha_\mathrm{\Theta}\vartheta - \alpha_\mathrm{c}c - \alpha_{\mathrm{\Theta}\mathrm{c}}\vartheta c.
\end{equation}
However, it has to be considered that the thermal expansion is essentially reduced when the temperature falls below the glass transition temperature. This is done by an extension proposed by \cite{leistnerdiss2022}, drawing on the logarithmic interpolation concept by \cite{kreisselmeiersteinhauser1979} to continuously combine two linear functions. The thermal deformation within the glassy state is indicated by an additional parameter $\alpha_{\mathrm{\Theta}\mathrm{G}}$. As a result, the thermo-chemical deformations are modeled as
\begin{equation}
	\label{eq:thermExp_chemShrink}
	J_{\mathrm{\Theta}\mathrm{c}}(\mathrm{\Theta},c) = 
    d\ln\Bigl(\exp\bigl( \alpha_\mathrm{\Theta}\vartheta/d \bigr) + \exp\bigl( (\alpha_{\mathrm{\Theta}\mathrm{G}}\vartheta + (\alpha_\mathrm{\Theta}-\alpha_{\mathrm{\Theta}\mathrm{G}}) (\mathrm{\Theta}_\mathrm{G}(c)-\mathrm{\Theta}_\mathrm{R}))/d \bigr)\Bigr)
    -\alpha_\mathrm{c}c - \alpha_{\mathrm{\Theta}c}\vartheta c + 1,
\end{equation}
where $d = \num[round-precision=0]{1.e-4}$ is a dimensionless curvature parameter that controls the smoothness of the transition between the two functions in the natural logarithm. The relatively small value of $d$ in Eq.~\eqref{eq:thermExp_chemShrink} leads to an abrupt change. The material parameters are identified based on density measurements using multiple steps, which are described in Sec.~\ref{sec:modelCalib}.

\paragraph{Specific heat capacity.} The specific heat capacity of thermosetting polymers typically increases with increasing temperature and exhibits a significant shift at the glass transition temperature. This shift is often considered by drawing on case distinction \cite{johnston1997,balversetal2008}. However, this is not necessary as shown in \cite{leistnerhartmannablizziegmann2020} since a simpler continuous model is sufficient and avoids case distinction,
\begin{equation}
	\label{eq:cp}
	c_\mathrm{p}^\mathrm{rev}(\mathrm{\Theta},c) = a_1 + a_2\mathrm{\Theta} + (a_3 + a_4\mathrm{\Theta})\tanh(a_5(\mathrm{\Theta}-\mathrm{\Theta}_\mathrm{G}(c))).
\end{equation}
The two linear functions describing the specific heat capacity below and above the glass transition temperature are continuously joined by the hyperbolic tangent. The material parameters $a_1,a_2,a_3$, and $a_4$ stem from the linear functions, whereas $a_5$ models the slope in the transition. The parameters are identified using heat flux data from temperature-modulated DSC (TMDSC) measurements at three different pre-conditioned curing states, namely, $c=0$, $c=\num[round-precision=2]{0.52}$, and $c=1$. TMDSC is a type of DSC in which the prescribed temperature profile is superimposed with a sinusoidal function, in our case with an amplitude of $\pm\SI[round-precision=0]{1}{\celsius}$ and a period of $\SI[round-precision=0]{60}{\s}$. The temperature-modulated procedure is crucial when chemical reactions occur during the DSC analysis and allows for separating reversing and non-reversing processes, see \cite{lionyagimli2008} for further details. The reversing part represents the specific heat capacity $c_\mathrm{p}^\mathrm{rev}$, while the non-reversing part is attributed to non-reversible processes, here, the chemical curing reaction \cite{lionyagimli2008}.

\paragraph{Thermal conductivity.} Epoxy resins exhibit a temperature- and degree of cure-dependent thermal conductivity. Fully cured epoxy resin shows a constant thermal conductivity with respect to temperature. This contrasts with the significantly temperature-dependent thermal conductivity of the uncured epoxy resin, which is accounted for by means of the logarithmic interpolation concept \cite{kreisselmeiersteinhauser1979}. The transition between uncured and cured resin is modeled with a simple rule of mixture,
\begin{equation}
	\label{eq:kappa}
	\kappa_\mathrm{\Theta}(\mathrm{\Theta},c) = b_1c + \tilde{d} \ln \left(\frac12\left(\exp(\kappa_{\mathrm{\Theta}\mathrm{a}}/d) + \exp(\kappa_{\mathrm{\Theta}\mathrm{b}}/d)\right)\right)(1-c).
\end{equation}
The parameter $\tilde{d} = \SI{0.01}{\W\per\m\K}$ is a curvature parameter within the logarithmic interpolation concept, smoothly interpolating between the two functions $\kappa_{\mathrm{\Theta}\mathrm{a}} = b_2$ and $\kappa_{\mathrm{\Theta}\mathrm{b}} = b_3(\mathrm{\Theta}-\mathrm{\Theta}_\mathrm{R})/\mathrm{\Theta}_\mathrm{R} + b_4$. Accordingly, $b_1,b_2,b_3$, and $b_4$ are the sought material parameters. Since the thermal conductivity can not be directly measured, we employ $\kappa_\mathrm{\Theta}(\mathrm{\Theta},c) = a_\mathrm{\Theta}(\mathrm{\Theta},c)\rho(\mathrm{\Theta},c)c_\mathrm{p}^\mathrm{rev}(\mathrm{\Theta},c)$. Consequently, the thermal diffusivity $a_\mathrm{\Theta}(\mathrm{\Theta},c)$ is measured with laser-flash analysis (LFA). This information is then combined with the parameter sets $\GKap_{J_{\mathrm{\Theta}\!c}}$ and $\GKap_{c_\mathrm{p}^\mathrm{rev}}$ to compute the experimental values of the thermal conductivity, see Fig.~\ref{fig:matparScheme}.

\section{Model Calibration using Nonlinear Least-Squares}
\label{sec:modelCalib}

The model calibration is conducted in multiple steps using the nonlinear least-squares method. Accordingly, the material parameter uncertainties are estimated in a frequentist setting based on the asymptotic properties of the nonlinear least-squares estimator, which is described first. Then, the multi-step model calibration of the constitutive relations is presented. Finally, the results are discussed. Note that the uncertainty propagation during the multi-step calibration is covered in Sec.~\ref{sec:uq_calib}.

\subsection{Frequentist Approach}
\label{sec:frequentist}
The nonlinear least-squares method is well-established for model calibration in computational solid mechanics, see \cite{roemerhartmanntroegerantonwesselsflascheldelorenzis2025} for an extensive review. Within each calibration step, the point estimate
\begin{equation}
	\label{eq:matparIdent}
	\GKap^* = \text{arg\,min}_{\GKap} \frac12 \lvert\lvert \bm{r}(\GKap) \rvert\rvert^2
\end{equation}
is sought for the $n_\kappa$ material parameters $\GKap\in\mathbb{R}^{n_\kappa}$. The residual $\bm{r} = \bm{s}(\GKap) - \bm{d}$, $\bm{r}\in\mathbb{R}^{n_\mathrm{D}}$, represents the difference between model response $\bm{s}(\GKap)$, $\bm{s}\in\mathbb{R}^{n_\mathrm{D}}$, and experimental observations $\bm{d}\in\mathbb{R}^{n_\mathrm{D}}$. Here, $n_\mathrm{D}$ denotes the number of experimental data and all data arrays are assembled as column vectors. 

An important, yet often neglected, aspect is the quantification of uncertainties associated with the point estimate $\GKap^*$. These uncertainties can be assessed by utilizing the asymptotic properties of the nonlinear least-squares estimator. It can be shown that $\GKap^*$ converges in distribution to a normal distribution \cite{jennrich1969},
\begin{equation}
	\label{eq:asymptProp}
	\sqrt{n_\mathrm{D}}(\GKap^* - \GKap_0) \rightarrow \mathcal{N}\left(\bm{0},\bm{Q}^{-1}(\GKap_0)\bm{Z}(\GKap_0)\bm{Q}^{-1}(\GKap_0)\right)
\end{equation}
with true but unknown deterministic $\GKap_0$. The asymptotic covariance matrix has the general form
\begin{equation}
	\label{eq:asympCovar}
	\bm{C} = \frac{1}{n_\mathrm{D}}\bm{Q}^{-1}(\GKap_0)\bm{Z}(\GKap_0)\bm{Q}^{-1}(\GKap_0).
\end{equation}
Eq.~\eqref{eq:asympCovar} is also known as the Huber sandwich estimator \cite{huber1967}, which provides robust estimates when model assumptions are violated \cite{freedman2006}. However, we follow the classical assumptions in nonlinear least-squares and assess robustness with synthetic tests in Sec.~\ref{sec:uq_calib}. Accordingly, we assume independent and identically distributed Gaussian observation noise $\bm{e}\sim\mathcal{N}(\bm{0},\sigma^2\bm{I})$ with finite variance $\sigma^2$. Under these conditions, the matrices $\bm{Q}$ and $\bm{Z}$ are approximated as
\begin{equation}
	\label{eq:Q_Z}
	\bm{Q}(\GKap_0) \approx \frac{2}{n_\mathrm{D}}\bm{J}^T(\GKap^*)\bm{J}(\GKap^*)
	\quad \text{and} \quad 
	\bm{Z}(\GKap_0) \approx \frac{4}{n_\mathrm{D}}\sigma^2 \bm{J}^T(\GKap^*)\bm{J}(\GKap^*),
\end{equation}
which yields
\begin{equation}
	\label{eq:asympCovar_short}
	\bm{C} \approx \sigma^2 (\bm{J}^T(\GKap^*)\bm{J}(\GKap^*))^{-1}.
\end{equation}
Here, $\bm{J} = \mathrm{d}\bm{s}/\mathrm{d}\GKap$ represents the Jacobian. The Jacobian-based approximation of the asymptotic covariance matrix in Eq.~\eqref{eq:asympCovar_short} is utilized to estimate material parameter uncertainties after inferring $\GKap^*$ from the experimental observations. The uncertainties of individual material parameters are computed from the variances, $\mathrm{\Delta}\kappa_{\mathrm{NLS};i} = \sqrt{C_{ii}}$, $i=1,\ldots,n_\kappa$. For further details on the derivation of Eq.~\eqref{eq:asympCovar_short}, we refer the reader to \cite{roemerhartmanntroegerantonwesselsflascheldelorenzis2025} and the references therein. Note that the above equations can be extended to a two-step model calibration setting \cite{troegerroemerhartmann2024}. However, as our model calibrations involve more than two subsequent steps, we quantify the propagation of uncertainties using FOSM and the Monte Carlo method in this work.

\subsection{Multi-Step Model Calibration}
\label{sec:multStepCalib}
\begin{figure}
    \centering
    \includegraphics[width=0.5\linewidth]{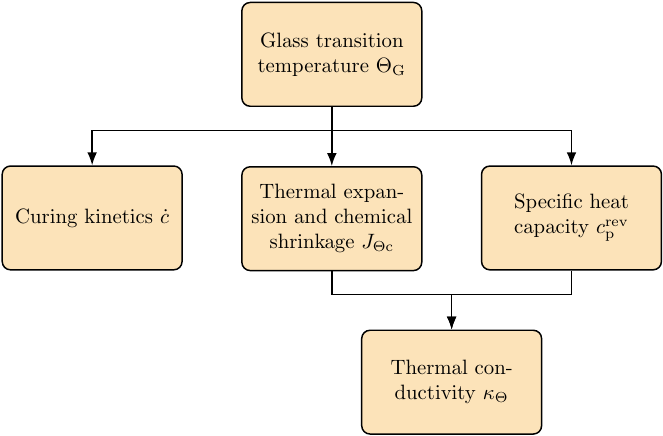}
    \caption{Steps during material parameter identification and dependencies}
    \label{fig:matparScheme}
\end{figure}

\begin{figure}[h!]
	\centering
	\subfloat
	[Glass transition temperature]
	{
	\includegraphics[width=0.35\textwidth]{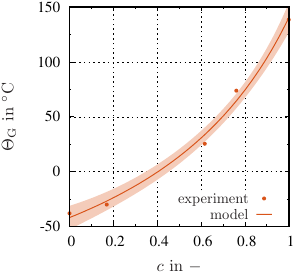}
	\label{fig:fit_thetaG}
	}
	\hspace{0.04\textwidth}
	\subfloat
	[Curing kinetics]
	{
	\includegraphics[width=0.35\textwidth]{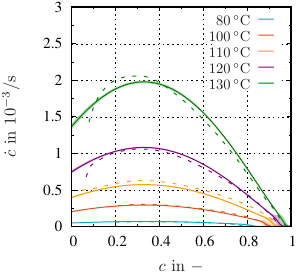}
	\label{fig:fit_kinetics}
	}
	\\
	\subfloat
	[Thermal expansion]
	{
	\includegraphics[width=0.35\textwidth]{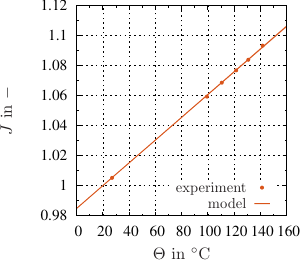}
	\label{fig:fit_thermExp}
	}
	\hspace{0.04\textwidth}
	\subfloat
	[Shrinkage]
	{
	\includegraphics[width=0.35\textwidth]{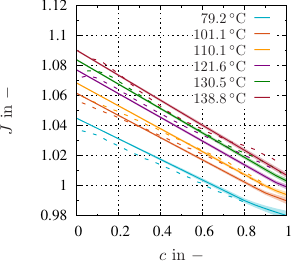}
	\label{fig:fit_shrink}
	}
	\\
	\subfloat
	[Specific heat capacity]
	{
	\includegraphics[width=0.35\textwidth]{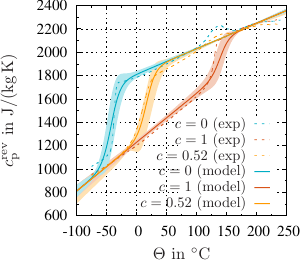}
	\label{fig:fit_cp}
	}
	\hspace{0.04\textwidth}
	\subfloat
	[Thermal conductivity]
	{
	\includegraphics[width=0.35\textwidth]{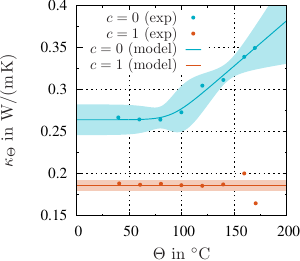}
	\label{fig:fit_kappa}
	}
	\caption{Experimental data and associated model response after identification of thermo-chemical material parameters with the nonlinear least-squares method. Shaded regions represent model uncertainty bands at a $\SI[round-precision=0]{95}{\percent}$ confidence level, estimated using the first-order second-moment method.}
	\label{fig:fit_uncert}
\end{figure}

The constitutive relations explained in Sec.~\ref{sec:constRelations} model the thermo-chemical curing and shrinkage behavior of epoxy resins using several material parameters, which have to be determined based on experimental data. We introduce a distinct material parameter set for each constitutive relation for clarity. These are compiled in Tab.~\ref{tab:matPar_exp}.
\begin{table}[ht]
	\centering
	\caption{Material parameter sets and experiments for constitutive relations in the thermo-chemical model}
	\label{tab:matPar_exp}
	\begin{tabular}{m{0.2\linewidth} m{0.28\linewidth} m{0.2\linewidth}}
	\toprule
	\textbf{model} & \textbf{material parameters} & \textbf{experiment} \\
	\midrule
	glass transition \newline temperature & $\GKap_{\mathrm{\Theta}_\mathrm{G}} = \lbrace r_\mathrm{f}, \mathrm{\Theta}_\mathrm{G0},\mathrm{\Theta}_\mathrm{G1} \rbrace^T$ & DSC \\
	curing kinetics & $\GKap_{\dot{c}} = \lbrace A, E, g, n, b_\mathrm{d} \rbrace^T$ & DSC \\
	thermal expansion and \newline chemical shrinkage & $\GKap_{J_{\mathrm{\Theta}\!c}} = \lbrace \alpha_\mathrm{\Theta}, \alpha_\mathrm{c}, \alpha_{\mathrm{\Theta}\mathrm{c}}, \alpha_{\mathrm{\Theta}\mathrm{G}} \rbrace^T$ & density measurements \\
	specific heat capacity & $\GKap_{c_\mathrm{p}^\mathrm{rev}} = \lbrace a_1, a_2, a_3, a_4, a_5 \rbrace^T$ & TMDSC \\
	thermal conductivity & $\GKap_{\kappa_\mathrm{\Theta}} = \lbrace b_1, b_2, b_3, b_4 \rbrace^T$ & LFA \\
	\bottomrule
	\end{tabular}
\end{table}
The constitutive relations are partly dependent on each other. For instance, the glass transition temperature $\mathrm{\Theta}_\mathrm{G}$ influences curing kinetics, chemical shrinkage, and specific heat capacity. Additionally, different experiments are performed to address the individual physical quantities and material parameters. Consequently, we perform the model calibration using multiple steps as illustrated in Fig.~\ref{fig:matparScheme}. 
Further details of the material parameter identification procedure are given in \cite{leistnerhartmannablizziegmann2020,leistnerdiss2022}, where especially the uniqueness of the material parameter estimates is investigated in detail. We briefly present the identification scheme to elaborate on the dependencies between the material parameter sets. Our focus lies on the uncertainty quantification since the uncertainty propagation was not considered in the aforementioned contributions. The experimental data employed during the multi-step calibration are visualized in Fig.~\ref{fig:fit_uncert}.

\paragraph{Glass transition temperature.} The DiBenedetto equation \eqref{eq:thetaG} yields the material parameter set $\GKap_{\mathrm{\Theta}_\mathrm{G}} = \lbrace r_\mathrm{f}, \mathrm{\Theta}_\mathrm{G0},\mathrm{\Theta}_\mathrm{G1} \rbrace^T$. The parameters are determined in a single step based on DSC data and are compiled in Tab.~\ref{tab:matpar}.
\begin{table}[ht!]
	\centering
	\caption{Material parameters of thermo-chemical curing model}
	\label{tab:matpar}
	\begin{tabular}{m{0.275\textwidth} l S S l}
	\toprule
	\multicolumn{2}{c}{\textbf{material parameter}} & $\GKap^*_\mathrm{NLS}$ & $\GKap^*_\mathrm{MC}$ & \multicolumn{1}{c}{\textbf{unit}} \\
	\cmidrule(lr){1-2} \cmidrule(lr){3-4} \cmidrule(lr){5-5}
	\multicolumn{5}{l}{\textit{glass transition temperature}} \\
	\midrule
	mobility ratio & $r_\mathrm{f}$ & \multicolumn{2}{c}{\num{4.4103e-01}} & -- \\
	glass transition \newline temperature (uncured) & $\mathrm{\Theta}_\mathrm{G0}$ & \multicolumn{2}{c}{\num{-4.1895966e01}} & \si{\celsius}\\
	glass transition \newline temperature (cured) & $\mathrm{\Theta}_\mathrm{G1}$ & \multicolumn{2}{c}{\num{1.403569e02}} & \si{\celsius} \\
	\midrule
	\multicolumn{5}{l}{\textit{curing kinetics}} \\
	\midrule
	pre-exponential factor & $A$ & \multicolumn{2}{c}{\num{5.01265e07}} & \si{\per\s} \\
	activation energy & $E$ & \multicolumn{2}{c}{\num{7.6594406e04}} & \si{\J\per\mol} \\
	factor & $g$ & \multicolumn{2}{c}{\num{3.517027e-01}} & -- \\
	order exponent & $n$ & \multicolumn{2}{c}{\num{1.4075975}} & -- \\
	diffusion factor & $b_\mathrm{d}$ & 4.83759159 & 9.629217 & \si{\K} \\
	\midrule
	\multicolumn{5}{l}{\textit{thermal expansion and chemical shrinkage}} \\
	\midrule
	thermal expansion coefficient & $\alpha_\mathrm{\Theta}$ & \multicolumn{2}{c}{\num{7.5878502e-04}} & \si{\per\K} \\
	chemical shrinkage coefficient & $\alpha_\mathrm{c}$ & 5.413657e-02 & 5.413676736e-02 & -- \\
	thermo-chemical coupling \newline coefficient & $\alpha_{\mathrm{\Theta}\mathrm{c}}$ & 2.462367e-04 & 2.4624335e-04 & \si{\per\K} \\
	glass transition coefficient & $\alpha_{\mathrm{\Theta}\mathrm{G}}$ & 6.96241016e-04 & 6.963413536e-04 & \si{\per\K} \\
	\midrule
	\multicolumn{5}{l}{\textit{specific heat capacity}} \\
	\midrule
	intercept parameter & $a_1$ & 1.52039194e03 & 1.520318899e03 & \si{\J\per\kg\per\K} \\
	slope parameter & $a_2$ & 3.18668457 & 3.17835309 &  \si{\J\per\kg\per\K\tothe{2}} \\
	intercept parameter & $a_3$ & 2.91465105e02 & 2.931239829e02 & \si{\J\per\kg\per\K} \\
	slope parameter & $a_4$ & -1.0017558 & -1.0015286 & \si{\J\per\kg\per\K\tothe{2}} \\
	shift parameter & $a_5$ & 6.688759e-02 & 6.25487597e-02 & \si{\per\K} \\
	\midrule
	\multicolumn{5}{l}{\textit{thermal conductivity}} \\
	\midrule
	parameter & $b_1$ & 1.8581859821e-01 & 1.85734377e-01 & \si{\W\per\m\per\K} \\
	parameter & $b_2$ & 2.7102864e-01 & 2.71171946e-01 & \si{\W\per\m\per\K} \\
	parameter & $b_3$ & 2.15493632e-02 & 2.15326497e-02 & \si{\W\per\m\per\K} \\
	parameter & $b_4$ & 1.94998300e-01 & 1.95157306e-01 & \si{\W\per\m\per\K} \\
	\bottomrule	
	\end{tabular}
\end{table}

\paragraph{Curing kinetics.} The simplified Kamal-Sourour model developed by Leistner \cite{leistnerhartmannablizziegmann2020} comprises the material parameter set $\GKap_{\dot{c}} = \lbrace A, E, g, n, b_\mathrm{d} \rbrace^T$. The material parameters are identified in two steps, since the diffusion-driven term becomes active only at an elevated degree of cure. First, the parameters of the chemically-driven part (i.e., $A$, $E$, $g$, and $n$) are identified. In a second step, the remaining diffusion parameter $b_\mathrm{d}$ is calibrated as soon as the diffusion-driven process becomes active, whereas the identified parameters of the chemically-driven part are fixed. 

\paragraph{Thermal expansion and chemical shrinkage.} Calibrating the thermo-chemical deformation behavior of epoxy resins presents greater challenges. The material parameters $\GKap_{J_{\mathrm{\Theta}\!c}} = \lbrace \alpha_\mathrm{\Theta}, \alpha_\mathrm{c}, \alpha_{\mathrm{\Theta}\mathrm{c}}, \alpha_{\mathrm{\Theta}\mathrm{G}} \rbrace^T$ are identified in multiple steps. First, the thermal expansion coefficient $\alpha_\mathrm{\Theta}$ of uncured resin is identified using the thermal expansion of a chemically stable uncured resin. Then, the parameters of chemical shrinkage $\alpha_\mathrm{c}$ and thermo-chemical coupling $\alpha_{\mathrm{\Theta}\mathrm{c}}$ are determined. To this end, density data of the curing process at various isothermal temperatures are employed. In a last step, the influence of the glass transition temperature is considered by identifying $\alpha_{\mathrm{\Theta}\mathrm{G}}$ based on density data for fully cured material at different temperatures.

\paragraph{Specific heat capacity.} The five material parameters of the set $\GKap_{c_\mathrm{p}^\mathrm{rev}}$ describe the temperature- and curing-dependent specific heat capacity. The parameters are calibrated leveraging heat flux data from a TMDSC measurement in a single step, whereas three degrees of cure ($c=0$, $c=0.52$, $c=1$) are considered.

\paragraph{Thermal conductivity.} Similar to the procedure for the specific heat capacity, the material parameters $\GKap_{\kappa_\mathrm{\Theta}}$ of the thermal conductivity are identified in a single step. The experimental values are obtained from the calibrated material parameters of shrinkage and specific heat capacity and LFA measurements according to the explanations in Sec.~\ref{sec:constRelations}. The data of an uncured and a fully cured epoxy resin are employed.

The material parameters of all constitutive relations are compiled in Tab.~\ref{tab:matpar} for clarity. The underlying experimental data and the model response with the identified thermo-chemical material parameters are illustrated in Fig.~\ref{fig:fit_uncert}. The parameters show very close agreement with earlier works \cite{leistnerhartmannablizziegmann2020,leistnerdiss2022}. However, the parameters of the chemically-driven curing kinetics term \eqref{eq:fc} are slightly different because we use an unweighted least-squares scheme, in contrast to the aforementioned works.

\begin{remark}
    In this work, we rigorously employ the multi-step calibration scheme shown in Fig.~\ref{fig:matparScheme}, using experimental data that target the individual material parameter sets and constitutive relations of interest. As explained below, uncertainties are propagated between the individual calibration steps using the first-order second-moment method. However, this approach might not fully resolve parameter correlations across the material parameter sets listed in Tab.~\ref{tab:matPar_exp}. Alternatively, a joint calibration of the different material parameter sets could be performed. This would require consideration of the different physical quantities involved as well as potentially significant differences in the amount of data available. Additionally, such a joint approach typically requires using the weighted nonlinear least-squares method. While it is possible to account for weights within the uncertainty quantification, choosing meaningful weights is challenging and, if done improperly, can lead to unreasonable parameter uncertainties. For these reasons, we use the multi-step calibration scheme and employ the unweighted nonlinear least-squares method. This approach enables us to determine reliable parameter uncertainty estimates using a Jacobian-based approximation of the covariance matrix, as outlined in Sec.~\ref{sec:frequentist}. 
\end{remark}

\section{Uncertainty Propagation during Inverse Analysis}
\label{sec:uq_calib}
Incorporating uncertainties essentially enhances the confidence in numerical simulation results. This consideration is also crucial during the multi-step model calibration since the calibrated parameters depend on each other, as shown in Fig.~\ref{fig:matparScheme}. Considering a set $\tilde{\GKap}^*\in\mathbb{R}^{n_{\tilde{\kappa}}}$ of $n_{\tilde{\kappa}}$ previously calibrated parameters, the solution \eqref{eq:matparIdent} of the nonlinear least-squares method can be rewritten for the multi-step case,
\begin{equation}
	\label{eq:matparIdent_multiStep}
	\GKap^*_\mathrm{NLS} = \text{arg\,min}_{\GKap} \frac12 \lvert\lvert \bm{r}(\GKap,\tilde{\GKap}^*) \rvert\rvert^2 = \text{arg\,min}_{\GKap} \frac12 \lvert\lvert \bm{s}(\GKap,\tilde{\GKap}^*) - \bm{d} \rvert\rvert^2.
\end{equation}
Apparently, the previously identified parameters $\tilde{\GKap}^*$ influence the model response $\bm{s}(\GKap,\tilde{\GKap}^*)$ and, hence, the obtained solution $\GKap^*$ of the current calibration step. 

\subsection{First-Order Second-Moment Method}
\label{sec:fosm_calib}
We investigate uncertainty propagation during model calibration using FOSM. Therein, the identified material parameters $\GKap^*$ are not affected by the uncertainty propagation because of the first-order expansion,
\begin{equation}
	\label{eq:expecFOSM_matpar}
	\GKap^* \approx \GKap^*_\mathrm{FOSM} = \GKap^*_\mathrm{NLS}.
\end{equation}
Consequently, the solution $\GKap^*_\mathrm{NLS}$ of the nonlinear least-squares scheme remains unchanged by uncertainty propagation. In contrast, the parameter covariance matrix reads
\begin{equation}
    \label{eq:covarFOSM_matpar}
    \bm{C}_{\GKap} = \underbrace{\left[\frac{\partial \GKap^*}{\partial \tilde{\GKap}^*}\right]\, \tilde{\bm{C}} \, \left[\frac{\partial \GKap^*}{\partial \tilde{\GKap}^*}\right]^T \!}_{\text{uncertainty propagation}} + \underbrace{\bm{C}\vphantom{\left[\frac{\partial \GKap^*}{\partial \tilde{\GKap}^*}\right]}}_{\text{noise}},
\end{equation}
where $\tilde{\bm{C}}$ and $\bm{C}$ represent the approximated covariance matrices based on asymptotic normality of the previous and current calibration step, respectively. The first part in Eq.~\eqref{eq:covarFOSM_matpar} considers uncertainty propagation from previously calibrated parameters, while the second term is the uncertainty due to measurement noise. The individual parameter variances are then
\begin{equation}
    \label{eq:varFOSM_matpar}
    (\delta\kappa_j)^2 \approx (\delta\kappa_{\mathrm{FOSM};j})^2 = \bm{e}_j^T\bm{C}_{\GKap}\bm{e}_j,
    \quad
    j = 1,\ldots,n_\kappa.
\end{equation}
The column vectors $\bm{e}_j\in\mathbb{R}^{n_\kappa}$ contain only zeros except a one at position $j$. A crucial aspect of applying the FOSM is the computation of the required derivatives $[\partial\GKap^*/\partial\tilde{\GKap}^*]$. Since no functional relationship between the parameters is available, the least-squares scheme is evaluated multiple times to approximate the derivatives using numerical difference quotients.

\subsection{Monte Carlo Method}
\label{sec:mc_calib}
While the FOSM is attractive due to its efficiency, it is important to keep in mind that it provides only a first-order approximation of the stochastic response and is limited to comparably small input parameter uncertainties. In contrast, stochastic approaches offer various options to quantify uncertainties during model calibration. Very popular are Bayesian methods, see \cite{rosicetal2013,rappelbeexhalenoelsbordas2020,roemerliuboel2022} and the literature cited therein. These techniques are also extended to track uncertainties during multi-step model calibration  \cite{roemerhartmanntroegerantonwesselsflascheldelorenzis2025,troegerroemerhartmann2024}. However, we use the Monte Carlo method as the reference to assess the FOSM results for uncertainty propagation during the calibration of nonlinear constitutive relations. 

The Monte Carlo approach simply draws a sample $\tilde{\GKap}^*_i$, $i=1,\ldots,n_\mathrm{MC}$, of the previously identified uncertain material parameters and repeatedly solves the nonlinear least-squares problem $\GKap^*_i = \text{arg\,min}_{\GKap} \frac12 \lvert\lvert \bm{r}(\GKap_i,\tilde{\GKap}^*_i) \rvert\rvert^2$, where we neglect the indication $_\mathrm{NLS}$ for brevity. In the multi-step calibration scheme, the uncertain input parameters are sampled as follows: material parameters $\tilde{\GKap}^*$ identified in the previous step (with their approximated covariance matrix $\tilde{\bm{C}}$) are sampled from a normal distribution $\tilde{\GKap}^*_i \sim \mathcal{N}(\tilde{\GKap}^*,\tilde{\bm{C}})$ according to Eq.~\eqref{eq:asymptProp}. In contrast, material parameters from steps earlier than the previous one are sampled from their empirical distributions $\tilde{\GKap}^*_i \sim \hat{\GKap}^*_\mathrm{MC}$ obtained after accounting for uncertainty propagation. After solving the nonlinear least-squares problem, i.e., identifying material parameters $\GKap^*_i$ and computing the approximation of their covariance matrix $\bm{C}_i$, for each of the $n_\mathrm{MC}$ samples, the moments of the random material parameters $\GKap^*$ are approximated by the sample mean
\begin{equation}
	\label{eq:expecMC_matpar}
	\GKap^* \approx \GKap^*_\mathrm{MC} = \frac{1}{n_\mathrm{MC}} \sum_{i=1}^{n_\mathrm{MC}} \GKap^*_i
\end{equation}
and covariance matrix
\begin{equation}
    \label{eq:covarMC_matpar_uncertProp}
    \hat{\bm{C}} = \frac{1}{n_\mathrm{MC}-1} \sum_{i=1}^{n_\mathrm{MC}} \left(\GKap^*_i - \GKap^*_\mathrm{MC} \right) \left(\GKap^*_i - \GKap^*_\mathrm{MC} \right)^T\!.
\end{equation}
Here, it is important to distinguish between the different sources of uncertainty being considered. The covariance matrix $\hat{\bm{C}}$ reflects only the uncertainty in the material parameters $\GKap^*_\mathrm{MC}$ arising from the propagation of uncertainty in the input parameters $\tilde{\GKap}^*$. Thus, similar to the first term in Eq.~\eqref{eq:covarFOSM_matpar}, $\hat{\bm{C}}$ depends on $\tilde{\bm{C}}$. However, the contribution of measurement noise to the overall material parameter uncertainty is not negligible and must also be taken into account. This contribution is calculated as the mean value of all individual Jacobian-based approximations of the covariance matrix
\begin{equation}
    \label{eq:covarMC_matpar_noise}
    \bm{C} = \frac{1}{n_\mathrm{MC}} \sum_{i=1}^{n_\mathrm{MC}} \bm{C}_i,
\end{equation}
where $\bm{C}_i$ is the approximated covariance matrix of material parameters $\GKap^*_i$ from the $i$th Monte Carlo run. Ultimately, the overall covariance matrix is obtained analogously to Eq.~\eqref{eq:covarFOSM_matpar}
\begin{equation}
    \label{eq:covarMC_matpar}
    \bm{C}_{\GKap} = \underbrace{\hat{\bm{C}}}_{\substack{\text{uncertainty}\\ \text{propagation}}} + \underbrace{\bm{C}}_{\text{noise}}.
\end{equation}
Thus, the Monte Carlo method yields the individual material parameter variances
\begin{equation}
    \label{eq:varMC_matpar}
    (\delta\kappa_j)^2 \approx (\delta\kappa_{\mathrm{MC};j})^2 = \bm{e}_j^T\bm{C}_{\GKap}\bm{e}_j,
    \quad
    j = 1,\ldots,n_\kappa.
\end{equation}
in the multi-step calibration scheme.

\subsection{Results and Discussion}
\label{sec:result_uq_calib}

\begin{figure}[ht]
	\centering
	\subfloat
	[Diffusion factor $b_\mathrm{d}$]
	{
	\includegraphics[width=0.3\textwidth]{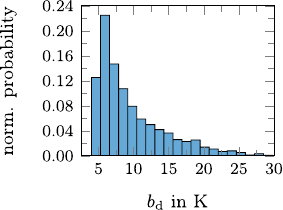}
	\label{fig:dist_bd}
	}
	\hspace{0.04\textwidth}
	\subfloat
	[Slope parameter $a_2$]
	{
	\includegraphics[width=0.3\textwidth]{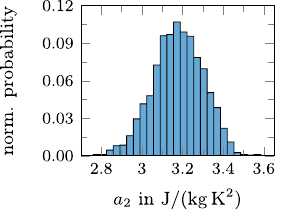}
	\label{fig:dist_a2}
	}
	\caption{Exemplary empirical probability distributions for identified material parameters using the Monte Carlo method}
	\label{fig:distMC}
\end{figure}

\paragraph{Material parameters.} The material parameters determined with the Monte Carlo approach are compiled in Tab.~\ref{tab:matpar} alongside the nonlinear least-squares results. For the Monte Carlo approach, $n_\mathrm{MC} = 2000$ samples were considered and the parameters were either sampled from a normal distribution according to Eq.~\eqref{eq:asymptProp} or from the empirical distribution during the multi-step calibration. Note that the parameters of the glass transition temperature \eqref{eq:thetaG}, the chemically-driven curing part \eqref{eq:fc}, and the thermal expansion coefficient are not affected by uncertainty propagation since all input parameters are deterministic. Consequently, the results of the nonlinear least-squares and Monte Carlo methods coincide for these parameters, i.e., $\GKap^*_\mathrm{NLS} = \GKap^*_\mathrm{MC}$. The majority of identified material parameters influenced by uncertain input parameters are in good agreement for the two methods. However, the diffusion factor $b_\mathrm{d}$ in the curing kinetics is a notable exception, as the approximated value strongly disagrees for both methods. The empirical probability distribution is illustrated in Fig.~\ref{fig:dist_bd}.
It becomes evident that the empirical distribution of the diffusion factor is clearly non-Gaussian, although all the random input parameters are Gaussian, causing a significant discrepancy between the parameter values. Note that the identified value of $b_\mathrm{d}$ within the nonlinear least-squares scheme coincides with the maximum probability in the empirical probability distribution. Thus, it can be concluded that the nonlinear least-squares method provides reasonable parameters even though considering only the nominal parameter value within the multi-step calibration of the thermo-chemical model.

\paragraph{Material parameter uncertainties.} The material parameter uncertainties $\mathrm{\Delta}\GKap_\mathrm{NLS}$ approximated within the nonlinear least-squares scheme, that is, without uncertainty propagation, and the uncertainties $\delta\GKap_\mathrm{FOSM}$ and $\delta\GKap_\mathrm{MC}$ considering uncertainty propagation in the multi-step calibration using FOSM and Monte Carlo method are given in Tab.~\ref{tab:uncert}.
\begin{table}[h!]
	\centering
	\caption{Material parameter uncertainties of thermo-chemical curing model. Values in brackets denote coefficients of variation computed with respect to $\GKap_\mathrm{NLS}^*$.}
	\label{tab:uncert}
	\begin{tabular}{m{0.23\textwidth} l S l S S l}
	\toprule
	\multicolumn{2}{c}{\textbf{material parameter}} & \multicolumn{2}{c}{$\mathrm{\Delta}\GKap_\mathrm{NLS}$} & $\delta\GKap_\mathrm{FOSM}$  & $\delta\GKap_\mathrm{MC}$ & \multicolumn{1}{c}{\textbf{unit}} \\
	\cmidrule(lr){1-2} \cmidrule(lr){3-4} \cmidrule(lr){5-6} \cmidrule(lr){7-7}
	\multicolumn{7}{l}{\textit{glass transition temperature}} \\
	\midrule
	mobility ratio & $r_\mathrm{f}$ & 6.73143e-02 & \scalebox{0.8}{(\SI{15.26}{\percent})} & \text{-} & \text{-} & -- \\
	glass transition \newline temperature \newline (uncured) & $\mathrm{\Theta}_\mathrm{G0}$ & 5.467009 & \scalebox{0.8}{(\SI{13.05}{\percent})} & \text{-} & \text{-} & \si{\celsius}\\
	glass transition \newline temperature \newline (cured) & $\mathrm{\Theta}_\mathrm{G1}$ & 6.7760428 & \scalebox{0.8}{(\SI{4.83}{\percent})} & \text{-} & \text{-} & \si{\celsius} \\
	\midrule
	\multicolumn{7}{l}{\textit{curing kinetics}} \\
	\midrule
	pre-exponential factor & $A$ & 5.2535796e06 & \scalebox{0.8}{(\SI{10.48}{\percent})} & \text{-} & \text{-} & \si{\per\s} \\
	activation energy & $E$ & 3.34195649e02 & \scalebox{0.8}{(\SI{0.44}{\percent})} & \text{-} & \text{-} & \si{\J\per\mol} \\
	factor & $g$ & 5.188022e-03 & \scalebox{0.8}{(\SI{1.48}{\percent})} & \text{-} & \text{-} & -- \\
	order exponent & $n$ & 1.50257429e-02 & \scalebox{0.8}{(\SI{1.07}{\percent})} & \text{-} & \text{-} & -- \\
	diffusion factor & $b_\mathrm{d}$ & 4.061173e-01 & \scalebox{0.8}{(\SI{8.40}{\percent})} & 2.3727446 & 5.013985338694305 & \si{\K} \\
	\midrule
	\multicolumn{7}{l}{\textit{thermal expansion and chemical shrinkage}} \\
	\midrule
	thermal expansion \newline coefficient & $\alpha_\mathrm{\Theta}$ & 2.55511e-06 & \scalebox{0.8}{(\SI{0.34}{\percent})} & \text{-} & \text{-} & \si{\per\K} \\
	chemical shrinkage \newline coefficient & $\alpha_\mathrm{c}$ & 2.07520867e-03 & \scalebox{0.8}{(\SI{3.83}{\percent})} & 2.07712342e-03 & 2.078665625123e-03 & -- \\
	thermo-chemical coupling \newline coefficient & $\alpha_{\mathrm{\Theta}\mathrm{c}}$ & 2.111280e-05 & \scalebox{0.8}{(\SI{8.57}{\percent})} & 2.1338366e-05 & 2.1352518971e-05 & \si{\per\K} \\
	glass transition coefficient & $\alpha_{\mathrm{\Theta}\mathrm{G}}$ & 8.8562847e-06 & \scalebox{0.8}{(\SI{1.27}{\percent})} & 1.98889720e-05 & 1.143870839492386e-05 & \si{\per\K} \\
	\midrule
	\multicolumn{7}{l}{\textit{specific heat capacity}} \\
	\midrule
	intercept parameter & $a_1$ & 3.25381325e-01 & \scalebox{0.8}{(\SI{0.02}{\percent})} & 1.169448254e01 & 1.2066592911237784e01 & \si{\J\per\kg\per\K} \\
	slope parameter & $a_2$ & 3.5159920e-03 & \scalebox{0.8}{(\SI{0.11}{\percent})} & 1.19997747e-01 & 1.23130673180194e-01 & \si{\J\per\kg\per\K\tothe{2}} \\
	intercept parameter & $a_3$ & 3.821120625e-01 & \scalebox{0.8}{(\SI{0.13}{\percent})} & 2.13707337 & 5.570925105814080 & \si{\J\per\kg\per\K} \\
	slope parameter & $a_4$ & 3.69965675e-03 & \scalebox{0.8}{(\SI{0.37}{\percent})} & 1.0754699e-01 & 1.04015677278713e-01 & \si{\J\per\kg\per\K\tothe{2}} \\
	shift parameter & $a_5$ & 4.33834523e-04 & \scalebox{0.8}{(\SI{0.65}{\percent})} & 2.16347325e-02 & 1.4236242725129e-02 & \si{\per\K} \\
	\midrule
	\multicolumn{7}{l}{\textit{thermal conductivity}} \\
	\midrule
	parameter & $b_1$ & 2.5950249e-03 & \scalebox{0.8}{(\SI{1.40}{\percent})} & 3.2863376e-03 & 3.513970548550e-03 & \si{\W\per\m\per\K} \\
	parameter & $b_2$ & 4.80503852e-03 & \scalebox{0.8}{(\SI{1.77}{\percent})} & 6.9033889e-03 & 5.351489235314e-03 & \si{\W\per\m\per\K} \\
	parameter & $b_3$ & 2.93888610e-03 & \scalebox{0.8}{(\SI{13.64}{\percent})} & 8.7184030e-03 & 3.044645524566e-03 & \si{\W\per\m\per\K} \\
	parameter & $b_4$ & 1.8288610e-02 & \scalebox{0.8}{(\SI{9.38}{\percent})} & 5.5583254e-02 & 1.8687217619833e-02 & \si{\W\per\m\per\K} \\
	\bottomrule
	\end{tabular}
\end{table}

The results yield several insights. First, it is noteworthy that the uncertainties associated with the estimated glass transition parameters are rather large due to the limited amount of available experimental data ($n_\mathrm{D} = 5$). This highlights the need to account for uncertainty propagation, given the dependencies between the constitutive relations depicted in Fig.~\ref{fig:matparScheme}. The necessity of uncertainty propagation is further underscored by the increased uncertainties observed for several parameters compared to the estimate $\mathrm{\Delta}\GKap_\mathrm{NLS}$ based on asymptotic normality. The latter is particularly pronounced for the material parameters of the specific heat capacity model \eqref{eq:cp}. Additionally, the uncertainties computed by FOSM and the Monte Carlo method are generally of the same order of magnitude and show very good agreement. However, larger discrepancies occur for parameters that do not exhibit a Gaussian distribution, such as the diffusion factor $b_\mathrm{d}$ (illustrated in Fig.~\ref{fig:dist_bd}) or the intercept parameter $a_3$. In contrast, uncertainty estimates are in good agreement for parameters with approximately Gaussian distribution, such as the slope parameter $a_2$ in Fig.~\ref{fig:dist_a2}. This observation is reasonable, as the FOSM does not account for any specific distribution information and provides only approximations of the mean value and variance.

In recent literature, FOSM is often regarded as being restricted to small input parameter uncertainties due to its first-order approximation of the response function, see \cite{kaminski2013,roemerbertschmulanischaeffer2022} for example. In the present study, the thermo-chemical model heavily depends on accurately identifying the parameters of the glass transition temperature, as shown in Fig.~\ref{fig:matparScheme}. However, the coefficients of variation of mobility ratio $r_\mathrm{f}$ and uncured glass transition temperature $\mathrm{\Theta}_\mathrm{G0}$ are approximately $0.15$ and $0.13$, respectively (computed with absolute value for $\mathrm{\Theta}_\mathrm{G0}$), which exceed the limit of $0.10$ suggested by \cite[pg. xii]{kaminski2013}. Additionally, we are dealing with highly nonlinear models, which at first glance might seem to preclude the use of FOSM. Nevertheless, the uncertainties computed using this method appear reasonable in the present work, as they show close agreement with those estimated using the Monte Carlo method. However, the limitations of the FOSM for approximating complex parameter distributions must be kept in mind.

\subsection{Coverage tests}
\label{sec:coverage}
In addition to the previous comparison between the FOSM and the Monte Carlo method, we further substantiate the robustness of our FOSM-based uncertainty quantification in the inverse setting by conducting coverage tests. While the previous results were obtained using real experimental data and the multi-step calibration scheme in Fig.~\ref{fig:matparScheme}, the coverage tests are performed on synthetic data generated from in-silico experiments since the true parameters must be known. Our objectives are two-fold: on the one hand, to evaluate the robustness of the Jacobian-based covariance matrix approximation in Eq.~\eqref{eq:asympCovar_short}, particularly when underlying assumptions, such as Gaussian observation noise, are violated. On the other hand, we assess the FOSM-based propagation of uncertainties in the multi-step calibration. To this end, we conduct three distinct studies, each addressing different aspects:
\begin{itemize}
    \item Case 1: uncertainty coverage for sparse data
    \item Case 2: uncertainty coverage for dense data and uncertainty propagation for a single parameter
    \item Case 3: uncertainty coverage for multiple parameters affected by uncertainty propagation
\end{itemize}
The specific settings of these studies are chosen to closely reflect the characteristics of our experimental data, ensuring the practical relevance of the coverage results to real calibration steps. Accordingly, \textit{case 2} and \textit{case 3} incorporate different types of observation noise, including not only Gaussian, but also uniform and heteroscedastic noise. Since the constitutive relations described in Sec.~\ref{sec:constRelations} are not necessarily limited to the specific epoxy resin considered here, we provide both the conditional coverage (as in classical frequentist coverage tests) and a more marginal coverage, averaged over the parameter domain. The latter is inspired by the procedure outlined in \cite{antontroegerwesselsroemerhenkeshartmann2025}.

\begin{figure}[ht]
    \centering
    \includegraphics[width=0.5\linewidth]{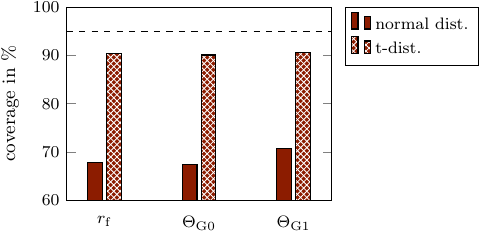}
    \caption{Marginal coverage of glass transition parameters $\GKap_{\mathrm{\Theta}_\mathrm{G}}$ ($n_\mathrm{D} = 5)$. The dashed line indicates the nominal confidence level of $\SI[round-precision=0]{95}{\percent}$.}
    \label{fig:coverage_thetaG}
\end{figure}

\paragraph{Case 1 - Sparse data.} The first case study addresses uncertainty coverage in the calibration of $\GKap_{\mathrm{\Theta}_\mathrm{G}}$, i.e., the parameters describing the glass transition temperature \eqref{eq:thetaG}. As illustrated in Fig.~\ref{fig:fit_thetaG}, the parameters are calibrated using very few data points, namely, $n_\mathrm{D} = 5$. For the coverage test, clean data is generated using the parameter values listed in Tab.~\ref{tab:matpar} for the material parameters $r_\mathrm{f}$, $ \mathrm{\Theta}_\mathrm{G0}$, and $\mathrm{\Theta}_\mathrm{G1}$. Subsequently, Gaussian random noise $\bm{e}_\mathrm{G} \sim \mathcal{N}(\bm{0},\sigma^2_{\mathrm{\Theta}_\mathrm{G}}\bm{I})$ with $\sigma_{\mathrm{\Theta}_\mathrm{G}} = \SI[round-precision=0]{4}{\celsius}$ is then added to the clean data. In total, $n_\mathrm{cov} = 1000$ independent repetitions of the model calibration are executed. The conditional coverage is evaluated with respect to the true material parameter values. For marginal coverage, the true material parameters are resampled in each repetition as $\GKap_{\mathrm{\Theta}_\mathrm{G};i} \sim \mathcal{N}(\GKap_{\mathrm{\Theta}_\mathrm{G}},\bm{C})$, $i = 1,\ldots,n_\mathrm{cov}$, and new clean data are generated accordingly. For simplicity, the covariance matrix $\bm{C}$ is taken as the Jacobian-based approximation obtained from the real experimental data. The results for the marginal coverage are illustrated in Fig.~\ref{fig:coverage_thetaG}.
Apparently, the empirical coverage is substantially below the nominal confidence level $\SI[round-precision=0]{95}{\percent}$ when confidence intervals are constructed assuming a normal distribution, i.e., a critical value of $\num[round-precision=2]{1.96}$. This shortfall is not surprising, given that any uncertainty quantification method is hardly robust when determining $n_\kappa = 3$ parameters using only $n_\mathrm{D} = 5$ data points. Specifically, the uncertainty estimates are based on assuming asymptotic normality (i.e., $n_\mathrm{D} \rightarrow \infty$), which is clearly violated here. However, when the sparse data is accounted for and the confidence intervals are constructed using the t-distribution, i.e., a critical value of approximately \num[round-precision=2]{4.30}, the coverage is essentially improved, reaching about $\SI[round-precision=0]{90}{\percent}$, see Fig.~\ref{fig:coverage_thetaG}. The quantitative results of the coverage test are summarized in Tab.~\ref{tab:coverage_thetaG}.
\sisetup{round-precision=1}
\begin{table}[ht]
    \centering
    \caption{Empirical coverage of glass transition parameters $\GKap_{\mathrm{\Theta}_\mathrm{G}}$ for $\SI[round-precision=0]{95}{\percent}$ nominal confidence level}
    \label{tab:coverage_thetaG}
    \begin{tabular}{m{2.8cm} c c c c c c}
    \toprule
     & \multicolumn{3}{c}{\textbf{normal distribution}} & \multicolumn{3}{c}{\textbf{t-distribution}} \\
     \cmidrule(lr){2-4} \cmidrule(lr){5-7}
     & $r_\mathrm{f}$ & $\mathrm{\Theta}_{\mathrm{G0}}$ & $\mathrm{\Theta}_\mathrm{G1}$ & $r_\mathrm{f}$ & $\mathrm{\Theta}_{\mathrm{G0}}$ & $\mathrm{\Theta}_\mathrm{G1}$ \\
    \midrule
    \multicolumn{3}{l}{$n_\mathrm{D} = 5$} \\
    \midrule
    conditional cov. & \SI{70.4}{\percent} & \SI{71.0}{\percent} & \SI{70.9}{\percent} & \SI{91.8}{\percent} & \SI{91.4}{\percent} & \SI{89.1}{\percent} \\
    marginal cov. & \SI{67.9}{\percent} & \SI{67.4}{\percent} & \SI{70.7}{\percent} & \SI{90.4}{\percent} & \SI{90.1}{\percent} & \SI{90.6}{\percent} \\
    \midrule
    \multicolumn{3}{l}{$n_\mathrm{D} = 50$} \\
    \midrule
    conditional cov. & \SI{93.0}{\percent} & \SI{93.4}{\percent} & \SI{95.0}{\percent} & \SI{93.8}{\percent} & \SI{93.9}{\percent} & \SI{95.2}{\percent} \\
    marginal cov. & \SI{95.2}{\percent} & \SI{93.7}{\percent} & \SI{94.8}{\percent} & \SI{95.5}{\percent} & \SI{94.2}{\percent} & \SI{95.2}{\percent} \\
    \bottomrule
    \end{tabular}
\end{table}
To study the effect of an increased amount of data, the coverage test with synthetic data was repeated with $n_\mathrm{D} = 50$. The corresponding empirical coverage values are reported in Tab.~\ref{tab:coverage_thetaG}. Notably, both the conditional and marginal coverage are now closer to the nominal confidence level. This indicates that the Jacobian-based covariance matrix approximation in Eq.~\eqref{eq:asympCovar_short} provides robust uncertainty estimates for the present case of nonlinear constitutive relations when a sufficient amount of data is available. Furthermore, the difference between the coverage results obtained from normal distribution and t-distribution (critical value is approximately \num[round-precision=2]{2.01} for $n_\mathrm{D} = 50$) is diminished.

\begin{figure}[ht]
	\centering
	\subfloat
	[Curing kinetics]
	{
	\includegraphics[width=0.32\textwidth]{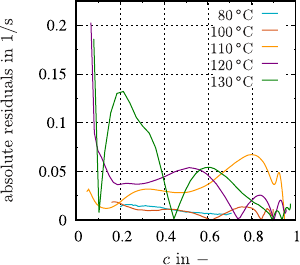}
	\label{fig:absRes_curingKin}
	}
	\hspace{0.04\textwidth}
	\subfloat
	[Specific heat capacity]
	{
	\includegraphics[width=0.32\textwidth]{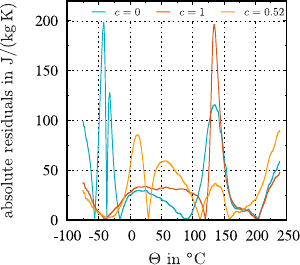}
	\label{fig:absRes_specHeat}
	}
	\caption{Absolute values of the residuals during calibration of (a) curing kinetics and (b) specific heat capacity using experimental data.}
	\label{fig:absRes}
\end{figure}

\paragraph{Case 2 - Dense data.} The second coverage test ($n_\mathrm{cov} = 1000$) focuses on the calibration of the curing kinetics parameters $\GKap_{\dot{c}}$. The rationale behind this choice is motivated by three key considerations: first, an accurate model of the curing kinetics is essential for producing reliable simulation results of curing materials. Second, dense data are present in contrast to the first study. Third, the parameter set $\GKap_{\dot{c}}$, compare Tab.~\ref{tab:matPar_exp}, allows for the simultaneous examination of uncertainty estimates for parameters that are not subjected to uncertainty propagation, as well as for the diffusion parameter $b_\mathrm{d}$, which is affected by propagating uncertainties. The covariance matrix approximation \eqref{eq:asympCovar_short} is based on the assumption of Gaussian observation noise, which might be violated in practice. The absolute values of the residuals observed during calibration of the curing kinetics using experimental data are visualized in Fig.~\ref{fig:absRes_curingKin}.
It is evident that the assumption of homoscedastic Gaussian observation noise is questionable for the calibration of the curing kinetics, since the residuals show non-constant variances. For the experiments conducted at lower temperatures, namely, $\SI[round-precision=0]{80}{\celsius}$ and $\SI[round-precision=0]{100}{\celsius}$, the homoscedastic noise assumption may be justified. However, this is clearly not the case at higher temperatures, where the curing kinetics display distinct transient behavior, compare Fig.~\ref{fig:fit_kinetics}. This finding is in agreement with the observations reported by \cite{jakobsenjensenandreasen2013}, wherein the authors noted that the signal-to-noise ratio decreases with increasing degree of cure, thereby reducing the precision of the enthalpy measurement. The latter forms the basis for calculating $\dot{c}$ and $c$, as described in Sec.~\ref{sec:constRelations}. Similarly, higher residuals are observed at the onset of the curing reaction at low degrees of cure and can be attributed to analogous reasons, shown in Fig.~\ref{fig:absRes_curingKin}.

\begin{figure}[ht]
	\centering
	\subfloat
	[]
	{
	\includegraphics[width=0.32\textwidth]{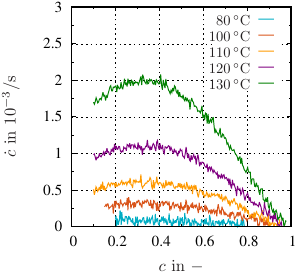}
	\label{fig:curingKin_Gaussian}
	}
	\subfloat
	[]
	{
	\includegraphics[width=0.32\textwidth]{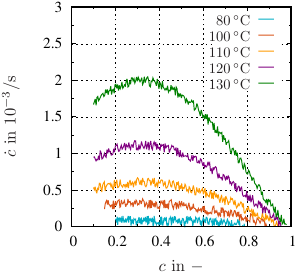}
	\label{fig:curingKin_uniform}
	}
	\subfloat
	[]
	{
	\includegraphics[width=0.32\textwidth]{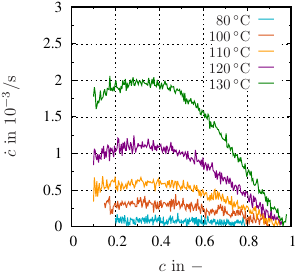}
	\label{fig:curingKin_het}
	}    
	\caption{Exemplary data from in-silico experiments of curing kinetics with different types of noise: (a) Gaussian noise $\bm{e}_\mathrm{G}$, (b) uniform noise $\bm{e}_\mathrm{U}$, (c) heteroscedastic Gaussian noise $\bm{e}_\mathrm{het}$.}
	\label{fig:curKin_noise}
\end{figure}

These observations motivate investigating the robustness of the covariance matrix approximation~\eqref{eq:asympCovar_short} in the presence of Gaussian, uniform, and heteroscedastic noise. The setup of our in-silico experiment is as follows, designed to closely mirror the real experimental data: $n_\mathrm{D} = 225$ data points are distributed equidistantly over the intervals $[0.2, 0.79]$, $[0.15, 0.9]$, $[0.1, 0.95]$, $[0.1, 0.96]$, and $[0.1, 0.98]$ for the degree of cure $c$ at different temperatures $\lbrace 80, 100, 110, 120, 130 \rbrace\,\si{\celsius}$. For each temperature, data up to \SI[round-precision=0]{90}{\percent} of the maximum degree of cure are used to calibrate the parameters $A$, $E$, $g$, and $n$ of the chemically-driven part of the curing kinetics~\eqref{eq:fc}. The remaining data are employed to determine the diffusion parameter $b_\mathrm{d}$ in Eq.~\eqref{eq:fd}, where uncertainty propagation from the glass transition temperature parameters and the parameter of the chemically-driven part must be considered. The clean data is augmented with three different types of noise:
\begin{itemize}
    \item Gaussian noise $\bm{e}_\mathrm{G} \sim \mathcal{N}(\bm{0},\sigma^2_{\dot{c}}\bm{I})$
    \item Uniform noise $\bm{e}_\mathrm{U} \sim \mathcal{U}(-\sigma_\mathrm{U}\bm{1}_{n_\mathrm{D}\times1},\sigma_\mathrm{U}\bm{1}_{n_\mathrm{D}\times1})$
    \item Heteroscedastic Gaussian noise $\bm{e}_\mathrm{het} \sim \mathcal{N}(\bm{0},\mathrm{diag}(\bm{\sigma}^2_\mathrm{het}))$
\end{itemize}
We choose $\sigma_{\dot{c}} = \SI[round-precision=0]{4.e-05}{\per\s}$, which corresponds to approximately \SI[round-precision=0]{2}{\percent} of the maximum curing rate $\dot{c}$ observed in our experiments, see Fig.~\ref{fig:fit_kinetics}. Accordingly, the relative noise level is significantly higher for in-silico experiments conducted at lower curing temperatures. To ensure comparability, the variances of uniform and Gaussian noise are matched, $\sigma_\mathrm{U} = \sqrt{3}\sigma_{\dot{c}}$. The individual standard deviations $\sigma_{\mathrm{het;}j}$, $j=1,\ldots,n_\mathrm{D}$, of the heteroscedastic Gaussian noise are modeled as a function of the degree of cure,
\begin{equation}
    \label{eq:sigHet_curKin}
    \sigma_{\mathrm{het;}j}(c) = \frac{k_1}{c_j + k_2} + k_3c_j,\qquad j=1,\ldots,n_\mathrm{D},
\end{equation}
where $k_1 = \num[round-precision=0]{1.e-05}$, $k_2 = \num[round-precision=0]{1.e-03}$, and $k_3 = \num[round-precision=1]{4.5e-05}$. These parameters are chosen so that the variance of the heteroscedastic noise is comparable to the other two noise types. Notably, the heteroscedastic Gaussian noise has a higher variance at lower curing degrees, which rapidly decays as observed in Fig.~\ref{fig:absRes_curingKin}. Then, the noise level linearly increases with the degree of cure $c$. Representative examples of data augmented with the three different types of noise are illustrated in Fig.~\ref{fig:curKin_noise}.
The coverage evaluation procedure for the parameters $A$, $E$, $g$, and $n$ is straightforward, as these parameters are not affected by uncertainty propagation. Hence, they serve to assess the effect of different noise types. Conditional coverage is determined for the true parameters in Tab.~\ref{tab:matpar}, which are used to generate clean data. The marginal coverage is obtained by sampling the true parameters in each repetition from a normal distribution using the uncertainties reported in Tab.~\ref{tab:uncert}. However, evaluating the coverage for the diffusion parameter $b_\mathrm{d}$ is more challenging, because it requires accounting for the uncertainty propagation of the glass transition temperature parameters $\GKap_{\mathrm{\Theta}_\mathrm{G}}$. This necessitates consideration of epistemic uncertainty in the coverage test and is realized by sampling both the glass transition parameters $\GKap_{\mathrm{\Theta}_\mathrm{G}}$ and the diffusion parameter $b_\mathrm{d}$ in each repetition to generate a new set of true parameters for coverage evaluation. Correspondingly, clean data are generated again in every repetition. Consistent with our experimental data, we use $n_{\mathrm{D};\mathrm{\Theta}_\mathrm{G}} = 5$ data points for the glass transition temperature data and apply Gaussian noise $\bm{e}_\mathrm{G} \sim \mathcal{N}(\bm{0},\sigma^2_{\mathrm{\Theta}_\mathrm{G}}\bm{I})$ as in the previous \textit{case 1}. All sampling steps utilize the identified parameters in Tab.~\ref{tab:matpar} and the respective uncertainties compiled in Tab.~\ref{tab:uncert}.

\begin{figure}[ht]
    \centering
    \includegraphics[width=0.85\linewidth]{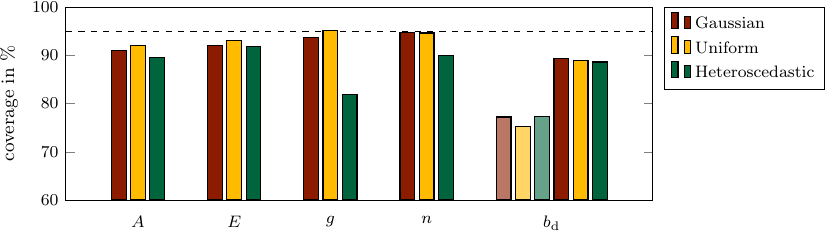}
    \caption{Marginal coverage of curing kinetics parameters $\GKap_{\dot{c}}$ ($n_{\mathrm{D};{\mathrm{\Theta}_\mathrm{G}}} = 5$). Lighter colors represent coverage without accounting for uncertainty propagation. The dashed line indicates the nominal confidence level of $\SI[round-precision=0]{95}{\percent}$.}
    \label{fig:coverage_curKin}
\end{figure}

The results for the marginal coverage are illustrated in Fig.~\ref{fig:coverage_curKin}.
The four parameters of the chemically-driven part exhibit good agreement between empirical coverage and nominal confidence level, although there is a slight tendency to state for the confidence intervals to be too narrow. This underestimation of parameter uncertainty is particularly pronounced for the factor $g$ in the presence of heteroscedastic noise, which is expected since the Jacobian-based covariance approximation in Eq.~\eqref{eq:asympCovar_short} assumes homoscedastic Gaussian noise. Furthermore, the necessity of taking uncertainty propagation into account is evident because the confidence intervals for the diffusion parameter $b_\mathrm{d}$ are too narrow and significantly underestimate the parameter uncertainty, see the light colored columns in Fig.~\ref{fig:coverage_curKin}. Employing the FOSM for uncertainty propagation improves the coverage for $b_\mathrm{d}$, although a slight underestimation relative to the nominal confidence level persists. The quantitative results are also compiled in Tab.~\ref{tab:coverage_curKin}.
\sisetup{round-precision=1}
\begin{table}[ht]
    \centering
    \caption{Empirical coverage of curing kinetics parameters $\GKap_{\dot{c}}$ for $\SI[round-precision=0]{95}{\percent}$ nominal confidence level. Values in brackets are obtained when neglecting uncertainty propagation.}
    \label{tab:coverage_curKin}
    \begin{tabular}{l m{2.8cm} c c c c c l}
    \toprule
     & & $A$ & $E$ & $g$ & $n$ & \multicolumn{2}{c}{$b_{\mathrm{d}}$} \\
    \midrule
    \multirow{6}{*}{\raisebox{-1.2cm}{\rotatebox[origin=c]{90}{\textbf{Gaussian}}}}
     & \multicolumn{7}{l}{$n_{\mathrm{D};{\mathrm{\Theta}_\mathrm{G}}} = 5$} \\
    \cmidrule{2-8}
     & conditional cov. & \SI{91.8}{\percent} & \SI{92.8}{\percent} & \SI{94.2}{\percent} & \SI{93.9}{\percent} & \SI{90.2}{\percent} & \scalebox{0.9}{(\SI{78.8}{\percent})} \\
     & marginal cov. &  \SI{91.0}{\percent} & \SI{92.1}{\percent} & \SI{93.7}{\percent} & \SI{94.7}{\percent} & \SI{89.3}{\percent} & \scalebox{0.9}{(\SI{77.2}{\percent})} \\
    \cmidrule{2-8}
     & \multicolumn{7}{l}{$n_{\mathrm{D};{\mathrm{\Theta}_\mathrm{G}}} = 50$} \\
    \cmidrule{2-8}
     & conditional cov. & \SI{91.4}{\percent} & \SI{92.9}{\percent} & \SI{94.7}{\percent} & \SI{95.3}{\percent} & \SI{90.1}{\percent} & \scalebox{0.9}{(\SI{80.6}{\percent})} \\
     & marginal cov. & \SI{91.5}{\percent} & \SI{93.2}{\percent} & \SI{94.4}{\percent} & \SI{95.1}{\percent} & \SI{89.9}{\percent} & \scalebox{0.9}{(\SI{81.0}{\percent})} \\
     \midrule
     \multirow{6}{*}{\raisebox{-1.0cm}{\rotatebox[origin=c]{90}{\textbf{Uniform}}}}
     & \multicolumn{7}{l}{$n_{\mathrm{D};{\mathrm{\Theta}_\mathrm{G}}} = 5$} \\
    \cmidrule{2-8}
     & conditional cov. & \SI{94.1}{\percent} & \SI{95.0}{\percent} & \SI{93.4}{\percent} & \SI{94.5}{\percent} & \SI{92.0}{\percent} & \scalebox{0.9}{(\SI{76.8}{\percent})} \\
     & marginal cov. & \SI{92.1}{\percent} & \SI{93.1}{\percent} & \SI{95.2}{\percent} & \SI{94.6}{\percent} & \SI{88.9}{\percent} & \scalebox{0.9}{(\SI{75.3}{\percent})} \\
    \cmidrule{2-8}
     & \multicolumn{7}{l}{$n_{\mathrm{D};{\mathrm{\Theta}_\mathrm{G}}} = 50$} \\
    \cmidrule{2-8}
     & conditional cov. & \SI{94.1}{\percent} & \SI{94.7}{\percent} & \SI{95.7}{\percent} & \SI{94.5}{\percent} & \SI{89.8}{\percent} & \scalebox{0.9}{(\SI{78.7}{\percent})} \\
     & marginal cov. & \SI{94.4}{\percent} & \SI{94.8}{\percent} & \SI{95.5}{\percent} & \SI{96.5}{\percent} & \SI{89.4}{\percent} & \scalebox{0.9}{(\SI{79.5}{\percent})} \\
     \midrule
     \multirow{6}{*}{\raisebox{-1.25cm}{\rotatebox[origin=c]{90}{\textbf{Heteroscedastic}}}}
     & \multicolumn{7}{l}{$n_{\mathrm{D};{\mathrm{\Theta}_\mathrm{G}}} = 5$} \\
    \cmidrule{2-8}
     & conditional cov. & \SI{92.3}{\percent} & \SI{92.7}{\percent} & \SI{82.8}{\percent} & \SI{90.3}{\percent} & \SI{93.3}{\percent} & \scalebox{0.9}{(\SI{82.1}{\percent})} \\
     & marginal cov. & \SI{89.6}{\percent} & \SI{91.9}{\percent} & \SI{81.8}{\percent} & \SI{90.0}{\percent} & \SI{88.6}{\percent} & \scalebox{0.9}{(\SI{77.3}{\percent})} \\
    \cmidrule{2-8}
     & \multicolumn{7}{l}{$n_{\mathrm{D};{\mathrm{\Theta}_\mathrm{G}}} = 50$} \\
    \cmidrule{2-8}
     & conditional cov. & \SI{88.6}{\percent} & \SI{91.3}{\percent} & \SI{83.0}{\percent} & \SI{92.1}{\percent} & \SI{92.1}{\percent} & \scalebox{0.9}{(\SI{84.5}{\percent})} \\
     & marginal cov. & \SI{91.0}{\percent} & \SI{93.2}{\percent} & \SI{82.6}{\percent} & \SI{92.3}{\percent} & \SI{89.9}{\percent} & \scalebox{0.9}{(\SI{82.5}{\percent})} \\
    \bottomrule
    \end{tabular}
\end{table}
The conditional and marginal coverage values are comparable across the different parameters and noise types, demonstrating that the obtained results hold in a certain neighborhood of the actual parameter values in Tab.~\ref{tab:matpar}. Additionally, there is no significant improvement in the coverage for the diffusion parameter $b_\mathrm{d}$ if the amount of data for the glass transition temperature $n_{\mathrm{D};\mathrm{\Theta}_\mathrm
G}$ is increased.

As a preliminary conclusion, the violation of the underlying assumptions of the covariance matrix approximation \eqref{eq:asympCovar_short}, for example, due to different noise or small amounts of data, can, but does not necessarily have to, negatively influence uncertainty coverage. The sensitivity to such violations varies among the material parameters. Additionally, although the FOSM may not fully resolve all nonlinear effects, the uncertainty estimates are improved at comparatively low computational costs.

\paragraph{Case 3 - Uncertainty propagation.} The final investigation of uncertainty coverage focuses on the specific heat capacity parameters $\GKap_{c_\mathrm{p}^\mathrm{rev}}$. In our real experiments, these parameters are calibrated using dense data ($n_\mathrm{D}$ on the order of $10^4$), but they are also subjected to uncertainty propagation from the glass transition temperature parameters $\GKap_{\mathrm{\Theta}_\mathrm{G}}$. Hence, in contrast to the previous study, five parameters are now affected by uncertainty propagation rather than just one. Similar to the reasoning in \textit{case 2}, the underlying experimental data from TMDSC analysis are very likely to violate the assumption of homoscedastic Gaussian noise, as indicated by the absolute values of the residuals shown in Fig.~\ref{fig:absRes_specHeat}. The pronounced peaks in the residuals in Fig.~\ref{fig:absRes_specHeat} can be attributed to the glass transition temperature at which a marked shift in the course of the specific heat capacity occurs, compare Fig.~\ref{fig:fit_cp}.

\begin{figure}[ht]
	\centering
	\subfloat
	[]
	{
	\includegraphics[width=0.32\textwidth]{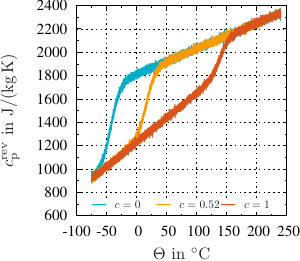}
	\label{fig:cp_Gaussian}
	}
	%
	%
	\subfloat
	[]
	{
	\includegraphics[width=0.32\textwidth]{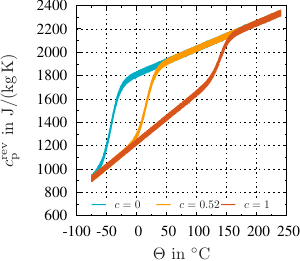}
	\label{fig:cp_uniform}
	}
    %
	%
	\subfloat
	[]
	{
	\includegraphics[width=0.32\textwidth]{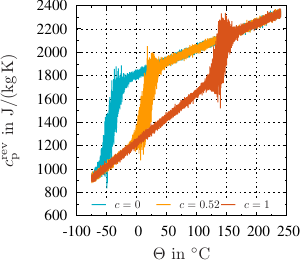}
	\label{fig:cp_het}
	}    
	\caption{Representative data from in-silico experiments of specific heat capacity with different types of noise: (a) Gaussian noise $\bm{e}_\mathrm{G}$, (b) uniform noise $\bm{e}_\mathrm{U}$, (c) heteroscedastic Gaussian noise $\bm{e}_\mathrm{het}$.}
	\label{fig:cp_noise}
\end{figure}

Again, the settings of our in-silico experiments are close to the real experiments. We distribute $n_\mathrm{D} = 17500$ data points evenly in the interval $[-75, 240]\,\si{\celsius}$ for the temperature $\mathrm{\Theta}$ and investigate three curing states $c = \lbrace 0.0, 0.52, 1.0 \rbrace$. The clean data is augmented with three types of noise:
\begin{itemize}
    \item Gaussian noise $\bm{e}_\mathrm{G} \sim \mathcal{N}(\bm{0},\sigma^2_{c_\mathrm{p}^\mathrm{rev}}\bm{I})$
    \item Uniform noise $\bm{e}_\mathrm{U} \sim \mathcal{U}(-\sigma_\mathrm{U}\bm{1}_{n_\mathrm{D}\times1},\sigma_\mathrm{U}\bm{1}_{n_\mathrm{D}\times1})$
    \item Heteroscedastic Gaussian noise $\bm{e}_\mathrm{het} \sim \mathcal{N}(\bm{0},\mathrm{diag}(\bm{\sigma}^2_\mathrm{het}))$
\end{itemize}
We choose $\sigma_{c_\mathrm{p}^\mathrm{rev}} = \SI{16.3}{\J\per\kg\per\K}$, which corresponds to \SI[round-precision=0]{1}{\percent} of the mean value of all experimental specific heat capacity data in our DSC analyses. Once again, Gaussian and uniform noise are defined to have the same variance, $\sigma_\mathrm{U} = \sqrt{3}\sigma_{c_\mathrm{p}^\mathrm{rev}}$. The heteroscedastic noise is modeled to account for the degree of cure-dependent glass transition temperature $\mathrm{\Theta}_\mathrm{G}(c)$, exhibiting the highest values at the glass transition temperature, while keeping a constant noise level elsewhere,
\begin{equation}
    \label{eq:sigHet_cp}
    \sigma_{\mathrm{het;}j}(\mathrm{\Theta},c) = \sigma_\mathrm{min} \left(1 + 7\exp\left(-\frac{(\mathrm{\Theta}_j - \mathrm{\Theta}_\mathrm{G}(c_j))^2}{2\omega^2} \right) \right), \quad j=1,\ldots,n_\mathrm{D}.
\end{equation}
The minimum noise level is selected as $\sigma_\mathrm{min} = \sigma_{c_\mathrm{p}^\mathrm{rev}}$ and $\omega = \SI[round-precision=0]{10}{\celsius}$ describes the width of region with higher noise level ($\sigma_\mathrm{max} = 8\sigma_\mathrm{min}$), which is centered around the glass transition temperature $\mathrm{\Theta}_\mathrm{G}$. Exemplary data of the specific heat capacity augmented with the three types of noise are visualized in Fig.~\ref{fig:cp_noise}.

The coverage tests are performed to address two main aspects. First, conditional coverage is evaluated for the identified parameters in Tab.~\ref{tab:matpar}, which are treated as true parameters to investigate the effect of different noise types. In this case, no uncertainties in the glass transition temperature parameters are propagated. Second, uncertainty propagation is taken into account by sampling both the glass transition temperature parameters, $\GKap_{\mathrm{\Theta}_\mathrm{G};i} \sim \mathcal{N}(\GKap_{\mathrm{\Theta}_\mathrm{G}},\bm{C}_{\mathrm{\Theta}_\mathrm{G}})$, $i = 1,\ldots,n_\mathrm{cov}$, and the specific heat capacity parameters, $\GKap_{c_\mathrm{p}^\mathrm{rev};i} \sim \mathcal{N}(\GKap_{c_\mathrm{p}^\mathrm{rev}},\bm{C}_{c_\mathrm{p}^\mathrm{rev}})$, in each repetition of the coverage test as new true parameters. The respective covariance matrices are set to the ones obtained during calibration from real experimental data for simplicity. Consequently, new clean data are generated in each repetition, where Gaussian noise $\bm{e}_{\mathrm{\Theta}_\mathrm{G}}$ is considered for the glass transition temperature data and the three different noise types stated above for the specific heat capacity data. Accordingly, the second investigation yields the marginal coverage and captures not only the effect of different noise types but also the uncertainty propagation, which is seen to be strongly pronounced according to the results in Tab.~\ref{tab:uncert}.

\begin{figure}[ht]
    \centering
    \includegraphics[width=0.8\linewidth]{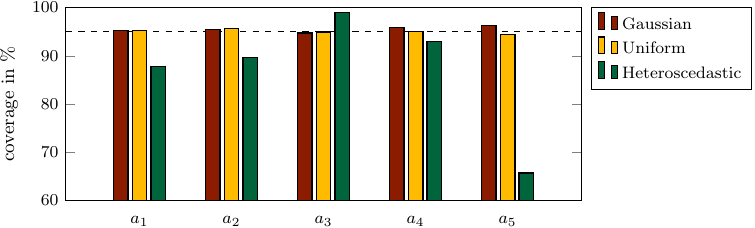}
    \caption{Conditional coverage of specific heat capacity parameters $\GKap_{c_\mathrm{p}^\mathrm{rev}}$ ($n_{\mathrm{D};{\mathrm{\Theta}_\mathrm{G}}} = 5)$ when no uncertain input parameters are considered and noise is the only source of uncertainty. The dashed line indicates the nominal confidence level of $\SI[round-precision=0]{95}{\percent}$.}
    \label{fig:coverage_cp_noise}
\end{figure}

The results for the conditional coverage are visualized in Fig.~\ref{fig:coverage_cp_noise}.
For both Gaussian and uniform noise, all five parameters show excellent coverage with respect to the nominal confidence level. This result is to be expected, given the large amount of data ($n_\mathrm{D} = \num[round-precision=2]{5.25e04}$), which ensures that the assumption of asymptotic normality is satisfied. Additionally, Gaussian observation noise aligns with the assumption during the derivations in Sec.~\ref{sec:frequentist}. In contrast, the modeled heteroscedastic Gaussian noise substantially impacts the coverage, especially for the shift parameter $a_5$. Within the model function~\eqref{eq:cp}, the shift parameter represents the glass transition temperature dependence and is therefore strongly affected by the heteroscedastic noise~\eqref{eq:sigHet_cp}. Note that the quantitative results are omitted for brevity.

\begin{figure}[ht]
    \centering
    \subfloat
	[$n_{\mathrm{D};{\mathrm{\Theta}_\mathrm{G}}} = 5$]
	{
	\includegraphics[width=0.975\textwidth]{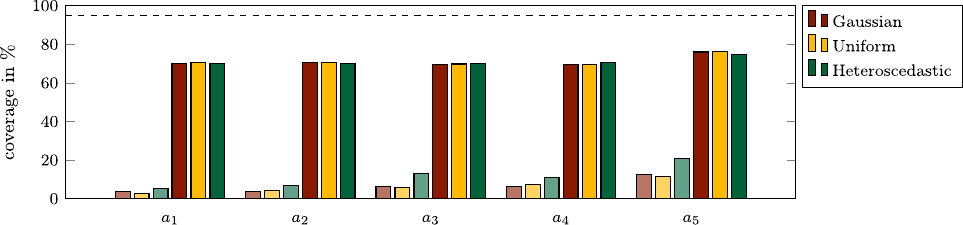}
	\label{fig:coverage_cp_nthetaG_5}
	}
    \\
    \subfloat
    [$n_{\mathrm{D};{\mathrm{\Theta}_\mathrm{G}}} = 50$]
    {
    \includegraphics[width=0.975\linewidth]{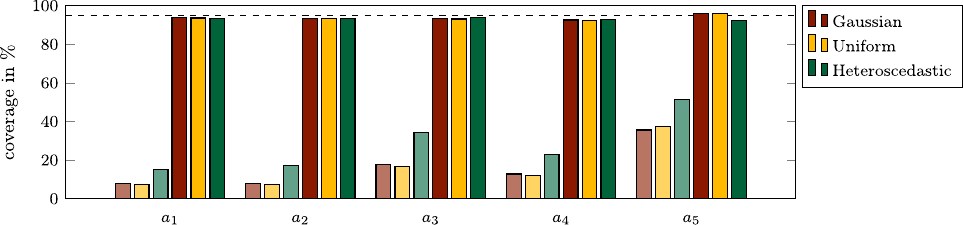}
    \label{fig:coverage_cp_nthetaG_50}
    }
    \caption{Marginal coverage of specific heat capacity parameters $\GKap_{c_\mathrm{p}^\mathrm{rev}}$ for different amounts of data $n_{\mathrm{D};{\mathrm{\Theta}_\mathrm{G}}}$ for calibrating glass transition temperature parameters. Lighter colors represent coverage without accounting for uncertainty propagation. The dashed line indicates the nominal confidence level of $\SI[round-precision=0]{95}{\percent}$.}
    \label{fig:coverage_cp}
\end{figure}

The empirical results for the marginal coverage obtained from the procedure explained above are visualized in Fig.~\ref{fig:coverage_cp}.
The results reveal two essential aspects. First, accounting for uncertainty propagation is crucial when estimating uncertainties in the specific heat capacity parameters, because the uncertain parameters of the glass transition temperature model~\eqref{eq:thetaG} significantly increase these parameter uncertainties. This finding is supported by the increasing parameter uncertainties reported in Tab.~\ref{tab:uncert} for both FOSM and the Monte Carlo method. Consequently, if uncertainty propagation is neglected, the empirical coverage falls short of the nominal confidence level, see Fig.~\ref{fig:coverage_cp_nthetaG_5}. However, it is important to note that the coverage is significantly below the nominal confidence level even when uncertainties are propagated using FOSM for sparse data $n_{\mathrm{D};{\mathrm{\Theta}_\mathrm{G}}} = 5$. The strong underestimation of uncertainty is not affected by the different types of measurement noise. The second essential aspect concerns the amount of data $n_{\mathrm{D};{\mathrm{\Theta}_\mathrm{G}}}$ used for calibrating the glass transition temperature parameters. Increasing $n_{\mathrm{D};{\mathrm{\Theta}_\mathrm{G}}}$ makes uncertainty propagation no less necessary; however, the empirical coverage very closely approaches the nominal confidence level, see Fig.~\ref{fig:coverage_cp_nthetaG_50}, even for heteroscedastic Gaussian noise. While this effect was not observed for the previous case of the diffusion parameter $b_\mathrm{d}$ in Tab.~\ref{tab:coverage_curKin}, it is unmissable in the results for the specific heat capacity parameters. The results are also quantitatively compiled in Tab.~\ref{tab:coverage_cp}.
\sisetup{round-precision=1}
\begin{table}[ht]
    \centering
    \caption{Empirical coverage of specific heat capacity parameters $\GKap_{c_\mathrm{p}^\mathrm{rev}}$ for $\SI[round-precision=0]{95}{\percent}$ nominal confidence level. Values in brackets are obtained when neglecting uncertainty propagation.}
    \label{tab:coverage_cp}
    \begin{tabular}{l m{2.5cm} c c c c c}
    \toprule
     & & {$a_1$} & {$a_2$} & {$a_3$} & {$a_4$} & {$a_5$} \\
    \midrule
    \multirow{6}{*}{\raisebox{-1.2cm}{\rotatebox[origin=c]{90}{\textbf{Gaussian}}}}
     & \multicolumn{6}{l}{$n_{\mathrm{D};{\mathrm{\Theta}_\mathrm{G}}} = 5$} \\
    \cmidrule{2-7}
     & marginal cov. & \SI{70.2}{\percent} & \SI{70.4}{\percent} & \SI{69.6}{\percent} & \SI{69.6}{\percent} & \SI{76.0}{\percent} \\
     & & \scalebox{0.9}{(\SI{3.6}{\percent})} & \scalebox{0.9}{(\SI{3.7}{\percent})} & \scalebox{0.9}{(\SI{6.4}{\percent})} & \scalebox{0.9}{(\SI{6.5}{\percent})} & \scalebox{0.9}{(\SI{12.5}{\percent})} \\
    \cmidrule{2-7}
     & \multicolumn{6}{l}{$n_{\mathrm{D};{\mathrm{\Theta}_\mathrm{G}}} = 50$} \\
    \cmidrule{2-7}
     & marginal cov. & \SI{93.7}{\percent} & \SI{93.2}{\percent} & \SI{93.3}{\percent} & \SI{92.6}{\percent} & \SI{96.0}{\percent} \\
     & & \scalebox{0.9}{(\SI{8.0}{\percent})} & \scalebox{0.9}{(\SI{7.7}{\percent})} & \scalebox{0.9}{(\SI{17.9}{\percent})} & \scalebox{0.9}{(\SI{12.8}{\percent})} & \scalebox{0.9}{(\SI{35.6}{\percent})} \\
    \midrule
    \multirow{6}{*}{\raisebox{-1.0cm}{\rotatebox[origin=c]{90}{\textbf{Uniform}}}}
     & \multicolumn{6}{l}{$n_{\mathrm{D};{\mathrm{\Theta}_\mathrm{G}}} = 5$} \\
    \cmidrule{2-7}
     & marginal cov. & \SI{70.4}{\percent} & \SI{70.6}{\percent} & \SI{69.8}{\percent} & \SI{69.4}{\percent} & \SI{76.3}{\percent} \\
     & & \scalebox{0.9}{(\SI{2.8}{\percent})} & \scalebox{0.9}{(\SI{4.1}{\percent})} & \scalebox{0.9}{(\SI{5.9}{\percent})} & \scalebox{0.9}{(\SI{7.4}{\percent})} & \scalebox{0.9}{(\SI{11.7}{\percent})} \\     
    \cmidrule{2-7}
     & \multicolumn{6}{l}{$n_{\mathrm{D};{\mathrm{\Theta}_\mathrm{G}}} = 50$} \\
    \cmidrule{2-7}
     & marginal cov. & \SI{93.6}{\percent} & \SI{93.2}{\percent} & \SI{93.1}{\percent} & \SI{92.5}{\percent} & \SI{96.1}{\percent} \\
     & & \scalebox{0.9}{(\SI{7.4}{\percent})} & \scalebox{0.9}{(\SI{7.4}{\percent})} & \scalebox{0.9}{(\SI{16.6}{\percent})} & \scalebox{0.9}{(\SI{11.9}{\percent})} & \scalebox{0.9}{(\SI{37.4}{\percent})} \\
    \midrule
    \multirow{6}{*}{\raisebox{-1.25cm}{\rotatebox[origin=c]{90}{\textbf{Heteroscedastic}}}}
     & \multicolumn{6}{l}{$n_{\mathrm{D};{\mathrm{\Theta}_\mathrm{G}}} = 5$} \\
    \cmidrule{2-7}
     & marginal cov. & \SI{70.0}{\percent} & \SI{70.2}{\percent} & \SI{70.1}{\percent} & \SI{70.7}{\percent} & \SI{74.5}{\percent} \\
     & & \scalebox{0.9}{(\SI{5.4}{\percent})} & \scalebox{0.9}{(\SI{6.7}{\percent})} & \scalebox{0.9}{(\SI{13.0}{\percent})} & \scalebox{0.9}{(\SI{11.2}{\percent})} & \scalebox{0.9}{(\SI{20.8}{\percent})} \\
    \cmidrule{2-7}
     & \multicolumn{6}{l}{$n_{\mathrm{D};{\mathrm{\Theta}_\mathrm{G}}} = 50$} \\
    \cmidrule{2-7}
     & marginal cov. & \SI{93.5}{\percent} & \SI{93.4}{\percent} & \SI{93.8}{\percent} & \SI{93.0}{\percent} & \SI{92.3}{\percent} \\
     & & \scalebox{0.9}{(\SI{15.0}{\percent})} & \scalebox{0.9}{(\SI{17.1}{\percent})} & \scalebox{0.9}{(\SI{34.1}{\percent})} & \scalebox{0.9}{(\SI{23.1}{\percent})} & \scalebox{0.9}{(\SI{51.4}{\percent})} \\
    \bottomrule
    \end{tabular}
\end{table}

In conclusion, this third coverage study of the specific heat capacity parameters demonstrates that uncertainty propagation is not only crucial for reliable uncertainty estimates, but also that the FOSM approach yields reliable results for nonlinear constitutive relations. Furthermore, it turned out that uncertainty estimates based on sparse data can have a substantial impact on the following model calibration steps and the associated uncertainty estimates. Whether such issues are significant in a given context can be assessed through in-silico experiments and coverage tests, as performed here, or through other methods such as sensitivity analysis.

\section{Uncertainty Propagation for Model Response}
\label{sec:uq_model}
Quantifying the effect of uncertain material parameters is important not only during multi-step model calibration but also, and perhaps more critically, when performing numerical simulations using these uncertain material parameters, as they inevitably lead to uncertain simulation results. In this work, we focus primarily on material parameter uncertainties, but we also compare these results with those obtained under uncertain boundary conditions. For related studies addressing various sources of uncertainty using FOSM, refer to \cite{dileephartmannetal2022,troegersartortigarhuomduesterhartmann2024,troegerkulozikhartmann2025}. 
We begin by briefly stating the problem formulation for transient thermal analysis of curing processes. Then, we proceed with short explanations of FOSM and the Monte Carlo method for quantifying model response uncertainties, with the latter serving again as a reference. Finally, we assess the uncertainty of numerical simulation results by studying the numerical example of a complex curing process.

\subsection{Transient Thermal Analysis of Curing Processes}
\label{sec:thermalCuring}
In a purely thermal setting, the balance of energy leads to the heat conduction equation
\begin{equation}
	\label{eq:heatPDE}
	c_\mathrm{p}^\mathrm{rev}(\mathrm{\Theta},c)\dot{\mathrm{\Theta}}(\vec{x},t) = - \frac{1}{\rho}\mathrm{div}\,\vec{q}(\vec{x},t) + h_\mathrm{c}\dot{c}(\mathrm{\Theta},c).
\end{equation}
Fourier's model is used for the heat flux vector $\vec{q}(\vec{x},t) = -\kappa_\mathrm{\Theta} \, \mathrm{grad}\,\mathrm{\Theta}(\vec{x},t)$. Special attention has to be paid to the second term on the right-hand side of the partial differential equation \eqref{eq:heatPDE}. The exothermic curing reaction is modeled as a volume-distributed heat source, see \cite{leistnerdiss2022} for details. In this work, we solve the heat equation accompanied with suitable initial and boundary conditions using finite elements for spatial discretization and higher-order diagonally-implicit Runge-Kutta methods for the time integration \cite{quinthartmannrothesabasteinhoff2011}. The use of higher-order time integration methods is necessary due to the inherent instability of the curing kinetics \eqref{eq:curingKin}. In contrast to typical thermal analyses, we are not only interested in computing the temperature field but also the degree of cure. Further details are beyond the scope of the present work and can be found in detail in \cite{leistnerhartmannablizziegmann2020}.

\subsection{First-Order Second-Moment Method}
\label{sec:fosm_model}
Adopting an abstract viewpoint on computational models, the quantities of interest, i.e., the model response, can be seen as a function of certain input parameters. In this study, we consider primarily uncertain material parameters $\GKap$ with mean $\GKap^*$ and standard deviation $\delta\GKap$ as uncertain input parameters. Consequently, we investigate the model $\mathcal{M}(\GKap)$. The following explanations can be readily extended to other sources of uncertainty, for example, uncertainty in boundary conditions, by substituting the mean and covariance matrix accordingly. Note that uncertainties arising from possible model simplifications are not taken into account. 

The FOSM approximation for expectation and variance of the model response reads
\begin{align}
	\label{eq:expecFOSM_model}
	\mathbb{E}[\mathcal{M}(\GKap)] &\approx \mu_{\mathrm{FOSM};\mathcal{M}} = \mathcal{M}(\GKap^*), \\
	\label{eq:varFOSM_model}
	\mathbb{V}[\mathcal{M}(\GKap)] &\approx \sigma^2_{\mathrm{FOSM};\mathcal{M}} = \sum_{i=1}^{n_\kappa} \sum_{j=1}^{n_\kappa} \mathcal{M},_i\!(\GKap^*) \mathcal{M},_j\!(\GKap^*) \mathrm{cov}[\kappa_i,\kappa_j].
\end{align}
Apparently, the model evaluated at the mean approximates the mean of the model response. Additionally, the variance is approximated using the model sensitivities and covariance information, which is readily available after material parameter identification. Note that the partial derivatives $\mathcal{M},_i = \partial \mathcal{M}(\GKap^*)/\partial \kappa_i$ are evaluated at the mean value $\GKap^*$. The FOSM can be efficiently implemented in an intrusive manner, see \cite{troegerkulozikhartmann2025}, or non-intrusively by drawing on finite difference approximations for the partial derivatives. 

The applications of the FOSM are two-fold. First, the uncertainty of the calibrated model response using experimental data is evaluated employing analytical derivatives of the constitutive relations in Sec.~\ref{sec:constRelations}. This is visualized as a shaded region in Fig.~\ref{fig:fit_uncert}, demonstrating acceptable model uncertainty due to parameter uncertainties $\delta\GKap_\mathrm{FOSM}$. Second, FOSM is applied to quantify the uncertainty of numerical simulation results later on. For the numerical example, the required derivatives are computed using finite differences as a non-intrusive approach. While the overarching goal is to compute output probability densities based on the input parameter densities, the FOSM is limited to Gaussian distributions that are entirely characterized by their expectation and variance. Additionally, the restriction to small input parameters due to the first-order approximation should be kept in mind, as discussed in \cite{roemerbertschmulanischaeffer2022}, which compares FOSM to the Monte Carlo method along with other methods.

\subsection{Monte Carlo Method}
\label{sec:mc_model}
Similar to the inverse setting, the Monte Carlo method requires drawing samples $\GKap_i \sim \GKap$, $i=1,\ldots,n_\mathrm{MC}$, of uncertain material parameters and performing repeated evaluations of the model $\mathcal{M}(\GKap_i)$ in a non-intrusive manner. As with the FOSM approach, the Monte Carlo method can be readily applied to uncertainties arising from boundary conditions or other sources by appropriately substituting the parameter mean and variance below. The expectation and variance of the model response are then approximated as
\begin{align}
	\label{eq:expecMC_model}
	\mathbb{E}[\mathcal{M}(\GKap)] &\approx \mu_{\mathrm{MC};\mathcal{M}} = \frac{1}{n_\mathrm{MC}} \sum_{i=1}^{n_\mathrm{MC}} \mathcal{M}(\GKap_i) \\
	\label{eq:varMC_model}
	\mathbb{V}[\mathcal{M}(\GKap)] &\approx \sigma^2_{\mathrm{MC};\mathcal{M}} = \frac{1}{n_\mathrm{MC}-1} \sum_{i=1}^{n_\mathrm{MC}} (\mathcal{M}(\GKap_i) - \mu_{\mathrm{MC};\mathcal{M}})^2
\end{align}
For numerical analyses using finite elements, the resulting systems of (non-)linear equations lead to multiple model functions. The extension is straightforward considering them as individual, but not necessarily independent model functions $\mathcal{M}_\alpha(\GKap)$, $\alpha = 1,\ldots,m$, with expectation $\mathbb{E}[\mathcal{M}_\alpha(\GKap)]$ and variance $\mathbb{V}[\mathcal{M}_\alpha(\GKap)]$. Note that we are not interested in computing the covariance between the different model outputs. Unfortunately, despite their robustness, the Monte Carlo method suffers from slow convergence and requires a large number of samples for high-dimensional inputs. Therefore, it is only used as a reference for comparison with the FOSM in this study.

\subsection{Numerical Example}
\label{sec:numExamp}
We investigate uncertainty quantification during the curing of an epoxy resin in an aluminum container, which serves as a validation example for studying curing processes \cite{leistnerhartmannablizziegmann2020}. First, the uncertainties associated with the various material parameters determined earlier are treated as uncertain input parameters. Subsequently, the effect of uncertain boundary conditions is assessed. In the former, a notable exception are thermal expansion and chemical shrinkage parameters, as their effects are indirectly considered in the thermal conductivity during calibration. However, a purely thermal simulation does not resolve shrinkage effects. Consequently, we use temperature and the degree of cure as model response quantities in our analysis. Uncertainty propagation is evaluated using the FOSM, with the Monte Carlo method serving as a reference. 

\subsubsection{Problem Setup.}
\begin{figure}[ht]
	\centering
	\includegraphics[width=0.42\textwidth]{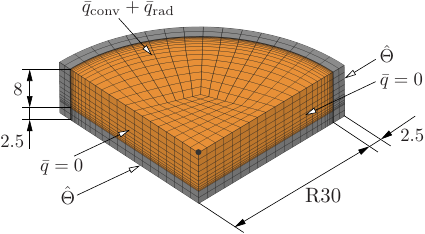}
	\caption{Geometry and boundary conditions for numerical investigation of curing of an epoxy resin (orange) in an aluminum container (gray). The circle indicates the node investigated in more detail.}
	\label{fig:meshBc}
\end{figure}

\begin{figure}[ht]
	\centering
	\includegraphics[width=0.65\textwidth]{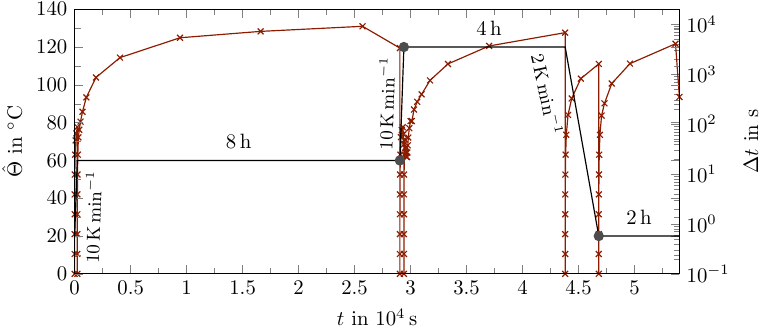}
	\caption{Piecewise linear temperature path during curing of epoxy resin (black) and time-adaptive time-step size (red). Gray circles indicate the time points evaluated in detail during uncertainty quantification.}
	\label{fig:curingPath_timeAdap}
\end{figure}

To reduce computational demands, we model only a quarter of the setup employing symmetry as shown in Fig.~\ref{fig:meshBc}.
The bottom and circumferential surface of the aluminum container are subjected to a Dirichlet boundary condition $\hat{\mathrm{\Theta}}(t)$, which is composed of piecewise linear temperature paths representing the temperature control during curing in an oven, see Fig.~\ref{fig:curingPath_timeAdap}.
The temperature path comprises three main phases -- first, a pre-curing or gelation phase (\SI[round-precision=0]{8}{\hour} at \SI[round-precision=0]{60}{\celsius}), then, a post-curing phase (\SI[round-precision=0]{4}{\hour} at \SI[round-precision=0]{120}{\celsius}), and, finally, a cooling phase (\SI[round-precision=0]{2}{\hour} at \SI[round-precision=0]{20}{\celsius}). The model is further subjected to adiabatic Neumann boundary conditions $\bar{q} = 0$ at the surfaces, where symmetry is employed. Additionally, heat exchange with the environment due to convection $\bar{q}_\mathrm{conv}$ and radiation $\bar{q}_\mathrm{rad}$ is considered at the top surface as a mixed boundary condition,
\begin{equation}
    \label{eq:qconv_qrad}
    \bar{q}_\mathrm{conv} = h(\mathrm{\Theta}-\mathrm{\Theta}_\infty)
    \quad\text{and}\quad 
    \bar{q}_\mathrm{rad} = \sigma \varepsilon (\mathrm{\Theta}^4 - \mathrm{\Theta}_\infty^4).
\end{equation}
We assume a convection coefficient $h = \SI[round-precision=0]{40}{\W\per\K\per\m\squared}$ and an emissivity $\varepsilon = \num[round-precision=1]{0.8}$. In Eq.~\eqref{eq:qconv_qrad}, $\sigma =\SI{5.67e-08}{\W\per\m\squared\per\K\tothe{4}}$ is the Stefan-Boltzmann constant and $\mathrm{\Theta}_\infty = \hat{\mathrm{\Theta}}$ the temperature of the surrounding fluid, which is set to the oven temperature $\hat{\mathrm{\Theta}}(t)$. The spatial discretization is carried out with 20-noded hexahedral elements and 8-noded surface elements to implement the mixed boundary condition of convective and radiative heat transfer. The aluminum container is modeled as isotropic with density $\rho = \SI[round-precision=1]{2.7}{\g\per\cm\cubed}$, specific heat capacity $c_\mathrm{p} = \SI{897}{\J\per\kg\per\K}$, and thermal conductivity $\kappa_\mathrm{\Theta} = \SI{235}{\W\per\m\per\K}$. Note that the material parameters of the aluminum container are treated as deterministic since the focus of the present work is on the epoxy resin.

Due to the inherent instability of the curing kinetics, a time-adaptive time integration procedure is highly beneficial for efficient computations to minimize the error evolution. Here, we employ embedded Runge-Kutta methods for the time integration, specifically, the second-order method by Ellsiepen with an embedded first-order method. The time-step size is controlled using a PI-controller, as detailed in \cite{gustafsson1994}. The time-step size evolution during this particular example is shown in Fig.~\ref{fig:curingPath_timeAdap}, clearly demonstrating the necessity for time-adaptivity with time-step sizes ranging from the initial value $\mathrm{\Delta} t = \SI[round-precision=0]{1.e-1}{\s}$ to \SI[round-precision=0]{8.5e3}{\s}.

\begin{figure}[h!]
	\centering
	\subfloat
	[$t = \SI{29040}{\s}$]
	{
	\includegraphics[width=0.4\textwidth]{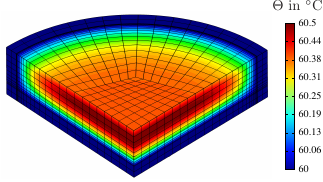}
	\label{fig:sim_Th_2}
	}
	\hspace{0.04\textwidth}
	\subfloat
	[$t = \SI{29040}{\s}$]
	{
	\includegraphics[width=0.4\textwidth]{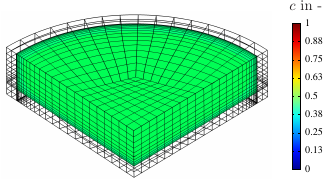}
	\label{fig:sim_c_2}
	}
	\\
	\subfloat
	[$t = \SI{29400}{\s}$]
	{
	\includegraphics[width=0.4\textwidth]{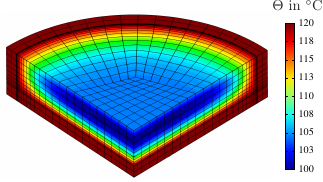}
	\label{fig:sim_Th_3}
	}
	\hspace{0.04\textwidth}
	\subfloat
	[$t = \SI{29400}{\s}$]
	{
	\includegraphics[width=0.4\textwidth]{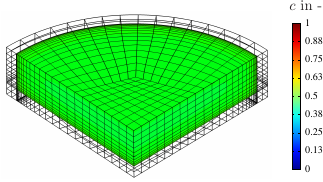}
	\label{fig:sim_c_3}
	}
	\\
	\subfloat
	[$t = \SI{46800}{\s}$]
	{
	\includegraphics[width=0.4\textwidth]{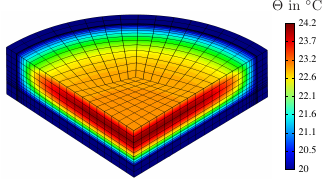}
	\label{fig:sim_Th_5}
	}
	\hspace{0.04\textwidth}
	\subfloat
	[$t = \SI{46800}{\s}$]
	{
	\includegraphics[width=0.4\textwidth]{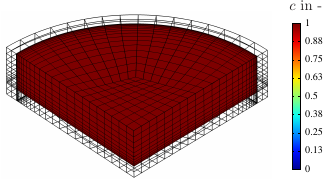}
	\label{fig:sim_c_5}
	}
	\caption{Simulation results of nodal temperatures (left) and curing degree (right) using the material parameters $\GKap^*_\mathrm{MC}$}
	\label{fig:sim_cylinder}
\end{figure}

The model response using the mean values of the parameter estimates $\GKap^*_\mathrm{MC}$ is illustrated in Fig.~\ref{fig:sim_cylinder} for three specific time points -- at the end of the gelation phase, the beginning of the post-curing phase, and the beginning of the cooling phase, compare Fig.~\ref{fig:curingPath_timeAdap}.
\sisetup{round-precision=0}
The temperature of the aluminum container closely follows the prescribed temperature profile $\hat{\mathrm{\Theta}}(t)$. In contrast, an allocation of heat is observed in the epoxy resin in the cured ($c\approx1$) state (Fig.~\ref{fig:sim_Th_5}), which is attributed to the comparably low thermal conductivity of the resin. The curing degree exhibits negligible spatial variability and increases rapidly during the transition from the pre-curing to the post-curing phase.

\subsubsection{Uncertainty Quantification.} 
The uncertainty quantification aims to evaluate the applicability of the FOSM by comparing its results with those obtained from the Monte Carlo method. We conduct three distinct case studies:
\begin{itemize}
    \item Case I: Effect of highly nonlinear constitutive relations and non-Gaussian distributions of uncertain material parameters
    \item Case II: Impact of increasing material parameter variance 
    \item Case III: Influence of uncertainties in the boundary conditions
\end{itemize}
\paragraph{Case I - Uncertain material parameters.} In the following, we use the parameters $\GKap^*_\mathrm{MC}$ (Tab.~\ref{tab:matpar}) and their respective uncertainties $\delta\GKap_\mathrm{MC}$ (Tab.~\ref{tab:uncert}). This choice allows us to evaluate the impact of clearly non-Gaussian distributions, as shown in Fig.~\ref{fig:dist_bd}, which are not explicitly considered in detail by the FOSM, thereby challenging the inherent limitations of the method. In the first study, we employ $n_\mathrm{MC} = 300$ samples for the Monte Carlo method. The material parameters $\GKap^*_i$ are sampled directly from a normal distribution $\GKap^*_i \sim \mathcal{N}(\GKap^*_\mathrm{MC},\bm{C})$, $i=1,\ldots,n_\mathrm{MC}$, if the parameter is not affected by uncertainty propagation during model calibration, compare Eq.~\eqref{eq:asymptProp}. Here, $\bm{C}$ represents the Jacobian-based approximation of the covariance matrix due to measurement noise, as introduced in Sec.~\ref{sec:frequentist}. In contrast, for material parameters influenced by propagating uncertainties during model calibration, the sampling procedure is more involved because the uncertainty contribution from noise is not negligible, compare Eq.~\eqref{eq:covarMC_matpar}. For these parameters, we proceed as follows:
\begin{enumerate}
    \item Sample material parameters $\check{\GKap}^*_i$ from the empirical distributions $\hat{\GKap}^*_\mathrm{MC}$ obtained from the Monte Carlo analysis in the inverse setting, $\check{\GKap}^*_i \sim \hat{\GKap}^*_\mathrm{MC}$, $i = 1,\ldots,n_\mathrm{MC}$.
    \item Determine the approximations of the covariance matrix $\bm{C}_i$ due to measurement noise, which are associated with $\check{\GKap}^*_i$.
    \item Sample material parameters $\GKap^*_i \sim \mathcal{N}(\check{\GKap}^*_i,\bm{C}_i)$, $i = 1,\ldots,n_\mathrm{MC}$ for the numerical simulation.
\end{enumerate}
At first glance, this sampling procedure may seem unnecessarily excessive. However, sampling the material parameters directly from their empirical distributions $\hat{\GKap}^*_\mathrm{MC}$, as obtained from the Monte Carlo analysis, only accounts for the covariance $\hat{\bm{C}}$ due to uncertainty propagation, i.e, only the first part in Eq.~\eqref{eq:covarMC_matpar}. This would result in underestimated output uncertainties due to the neglect of measurement noise. To avoid this, the associated approximated covariance matrix $\bm{C}_i$ is determined in step 2 and subsequently incorporated in step 3 by exploiting the asymptotic properties of the nonlinear least-squares estimator, Eq.~\eqref{eq:asymptProp}. In contrast to the Monte Carlo method, the FOSM does not rely on samples, as introduced in Sec.~\ref{sec:fosm_model}. The FOSM is evaluated using the mean and variance of the sampled material parameters.

\paragraph{Case II - Scaled material parameter variance.} While the first study compares FOSM and the Monte Carlo method for parameter uncertainties obtained during model calibration, the second study probes the inherent limitations of FOSM that arise from its first-order approximation of the stochastic response. In this second study, this is accomplished by scaling the variance of the considered material parameters. The variance scaling is applied to each of the $n_\kappa$ material parameters as follows:
\begin{enumerate}
    \item[] Input: material parameter samples $\GKap_j$ obtained as outlined for \textit{case I}
    \item Compute the mean of the sampled material parameter $\bar{\kappa}_j$ 
    \item Center the samples $\mathrm{\Delta}\GKap_j = \GKap_j - \bar{\kappa}_j \bm{1}_{n_\mathrm{MC}\times 1}$
    \item Scale the deviations $\mathrm{\Delta}\GKap^\mathrm{infl}_j = \sqrt{k}\mathrm{\Delta}\GKap_j$
    \item Construct the inflated empirical distribution $\GKap^\mathrm{infl}_j = \bar{\kappa}_j \bm{1}_{n_\mathrm{MC}\times 1} + \mathrm{\Delta}\GKap^\mathrm{infl}_j$    
\end{enumerate}
The procedure above is performed for each individual material parameter $j$, $j = 1,\ldots,n_\kappa$ and $\bm{1}_{n_\mathrm{MC}\times 1}$ denotes a column vector of ones with length $n_\mathrm{MC}$. Analogous to the first study, $n_\mathrm{MC} = 300$ samples are used for the Monte Carlo analysis in the second study. The individual material parameter samples $\GKap_j$ are obtained as explained for \textit{case I} above. Since the scaling procedure is applied to the empirical distributions, the approximated covariance matrices $\bm{C}$ arising from noise must be scaled separately as $k\bm{C}$ before sampling. After scaling, the inflated material parameter distributions $\GKap^\mathrm{infl}$ retain the previous mean values, but their variance is now scaled with $k$. For this second study, we use $k = 1,2,10$ to compare the FOSM results with those obtained from the Monte Carlo method. For simplicity, all material parameter variances are scaled simultaneously by the same factor $k$. The FOSM results are straightforward to compute using the same partial derivatives $\mathcal{M},_j$ for all scaling factors, since these are evaluated at the unchanged mean value. Accordingly, only the scaled variance of the material parameters has to be considered.

\paragraph{Case III - Uncertain boundary conditions.} In the third study, we treat the material parameters $\GKap = \GKap^*_\mathrm{MC}$ as deterministic and instead account for uncertainties in the boundary conditions. The different boundary conditions are illustrated in Fig.~\ref{fig:meshBc}. The adiabatic boundary condition $\bar{q} = 0$ results from modeling only a quarter of the setup and remains fixed, as there is no natural variability associated with it. In contrast, the prescribed temperature profile $\hat{\mathrm{\Theta}}(t)$, representing the curing process in an oven, is subjected to fluctuations. This piecewise-linear path, shown in Fig.~\ref{fig:curingPath_timeAdap}, consists of seven distinct temperatures $\hat{\mathrm{\Theta}} = \lbrace 20, 60, 60, 120, 120, 20, 20 \rbrace\si{\celsius}$, specified at the start and end of each segment. We consider these seven temperature values as uncertain and denote them as $\hat{\mathrm{\Theta}}_j$, $j=1,\ldots,7$. The individual temperature profiles between these points are assumed to remain linear, while the corresponding time points are kept fixed. The mean values are set equal to the original temperatures in the curing profile $\mu_{\mathrm{\Theta}_j} = \hat{\mathrm{\Theta}}_j$. The uncertainty is defined as \SI{10}{\percent} of the absolute temperature value, $\sigma_{\mathrm{\Theta}_j} = 0.1\mu_{\mathrm{\Theta}_j}$. Samples for the Monte Carlo method are drawn from a normal distribution $\hat{\mathrm{\Theta}}_{j;i} \sim \mathcal{N}(\mu_{\mathrm{\Theta}_j},\sigma^2_{\mathrm{\Theta}_j})$, $i=1,\ldots,n_\mathrm{MC}$. To adequately represent the variability in the Dirichlet boundary conditions $\hat{\mathrm{\Theta}}$, we choose $n_\mathrm{MC} = 150$.

As visualized in Fig.~\ref{fig:meshBc}, we model convective and radiative heat exchange on the top surface as a mixed boundary condition according to Eq.~\eqref{eq:qconv_qrad}. Although convection in an oven is generally expected to be comparably constant, the specific value of the convection coefficient $h$ is inherently uncertain, as it might not accurately reflect the true convective heat transfer. Similarly, the surface emissivity $\varepsilon$ is a modeling choice and is also subject to uncertainty. Therefore, both values are treated as uncertain parameters to evaluate the impact of uncertain parameters in the mixed boundary conditions on the nodal temperatures and degree of cure. The mean value of the convection coefficient is set to correspond to the value used in the original numerical simulations, $\mu_h = \SI{40}{\W\per\K\per\m\squared}$. Consistent with previous assumptions, we specify a $\SI{10}{\percent}$ uncertainty, $\sigma_h = 0.1\mu_h = \SI{4}{\W\per\K\per\m\squared}$. Since the convection coefficient should not be negative, we opt for the lognormal distribution $h_i \sim \mathrm{Log}\mathcal{N}(\mu_\mathrm{ln},\sigma^2_\mathrm{ln})$. The distribution parameters are calculated from the specified moments,
\begin{align}
    \label{eq:convCoeffDist_mu}
    \mu_\mathrm{ln} &= \ln \left(\frac{\mu_h}{h_0}\right) - \frac{1}{2} \ln\left( \frac{\sigma_h^2}{\mu_h^2} + 1 \right), \\
    \label{eq:convCoeffDist_sig}
    \sigma_\mathrm{ln} &= \sqrt{\ln\left( \frac{\sigma_h^2}{\mu_h^2} + 1 \right)}.
\end{align}
Note that $h_0 = \SI{1}{\W\per\K\per\m\squared}$ is introduced solely to render the argument of the natural logarithm dimensionless. 

\begin{figure}[ht]
	\centering
	\subfloat
	[Convection coefficient $h$]
	{
	\includegraphics[width=0.3\textwidth]{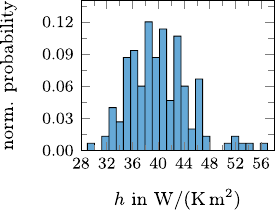}
	\label{fig:dist_convCoeff}
	}
	\hspace{0.04\textwidth}
	\subfloat
	[Emissivity $\varepsilon$]
	{
	\includegraphics[width=0.3\textwidth]{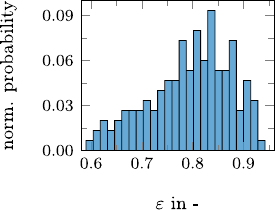}
	\label{fig:dist_emissiv}
	}
	\caption{Empirical probability distributions of randomly sampled boundary condition values used in the uncertainty quantification study}
	\label{fig:dist_uncertain_bc}
\end{figure}

We proceed similarly for the surface emissivity, selecting the mean value $\mu_\varepsilon = \num[round-precision=1]{0.8}$ and an uncertainty $\sigma_\varepsilon = 0.1\mu_\varepsilon = \num[round-precision=2]{0.08}$. However, it must be considered that the surface emissivity can only take values within $[0,1]$. Hence, we choose the beta distribution and sample the emissivities for the Monte Carlo analysis $\varepsilon_i \sim \mathrm{Beta}(\alpha,\beta)$. As before, the distribution parameters $\alpha$ and $\beta$ are determined based on the specified moments,
\begin{align}
    \label{eq:emissivDist_alpha}
    \alpha &= \left(\frac{\mu_\varepsilon\left(1-\mu_\varepsilon\right)}{\sigma^2_\varepsilon} - 1 \right) \mu_\varepsilon, \\
    \label{eq:emissivDist_beta}
    \beta &= \left(\frac{\mu_\varepsilon\left(1-\mu_\varepsilon\right)}{\sigma^2_\varepsilon} - 1 \right) \left(1 - \mu_\varepsilon\right).
\end{align}
The empirical distributions for both convection coefficient and emissivity are visualized in Fig.~\ref{fig:dist_uncertain_bc}.
All settings for investigating the impact of uncertain boundary conditions on the model response (nodal temperatures, degree of cure) are compiled in Tab.~\ref{tab:uncertBC}.
\begin{table}[ht]
    \centering
    \caption{Uncertain boundary conditions assumed in the third study (\textit{case III})}
    \label{tab:uncertBC}
    \begin{tabular}{m{4.2cm} c c c}
    \toprule
    \textbf{boundary condition} & $\mu$ & $\sigma$ & \textbf{distribution} \\
    \midrule
    Temperature $\hat{\mathrm{\Theta}}_j$, {\small{$j=1,\ldots,7$}} & $\lbrace 20, 60, 60, 120, 120, 20, 20 \rbrace\,\si{\celsius}$ & $\lbrace 2, 6, 6, 12, 12, 2, 2 \rbrace\,\si{\celsius}$ & $\mathcal{N}(\mu_{\mathrm{\Theta}_j},\sigma^2_{\mathrm{\Theta}_j})$ \\
    Convection coefficient $h$ & $\SI[per-mode=symbol]{40}{\W\per\K\per\m\squared}$ & $\SI[per-mode=symbol]{4}{\W\per\K\per\m\squared}$ & $\mathrm{Log}\mathcal{N}(\mu_\mathrm{ln},\sigma^2_\mathrm{ln})$ \\
    Emissivity $\varepsilon$ & \num[round-precision=1]{0.8} & \num[round-precision=2]{0.08} & $\mathrm{Beta}(\alpha,\beta)$ \\
    \bottomrule
    \end{tabular}
\end{table}
We investigate two settings: first, uncertain Dirichlet and mixed boundary conditions with $n_\mathrm{MC} = 150$, and subsequently, only uncertain mixed boundary conditions by keeping the temperatures constant and varying only the convection coefficient and emissivity with $n_\mathrm{MC} = 60$.

\subsubsection{Results and Discussion.} 
\begin{figure}[h!]
	\centering
	\subfloat
	[$t = \SI{29040}{\s}$]
	{
	\includegraphics[width=0.4\textwidth]{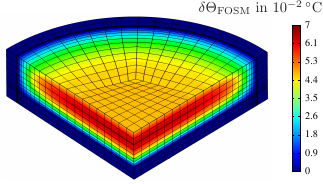}
	\label{fig:fosm_Th_2}
	}
	\hspace{0.04\textwidth}
	\subfloat
	[$t = \SI{29040}{\s}$]
	{
	\includegraphics[width=0.4\textwidth]{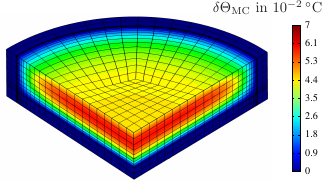}
	\label{fig:mc_Th_2}
	}
	\\
	\subfloat
	[$t = \SI{29400}{\s}$]
	{
	\includegraphics[width=0.4\textwidth]{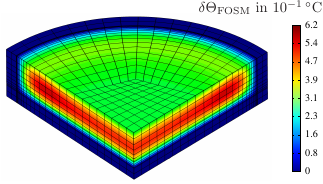}
	\label{fig:fosm_Th_3}
	}
	\hspace{0.04\textwidth}
	\subfloat
	[$t = \SI{29400}{\s}$]
	{
	\includegraphics[width=0.4\textwidth]{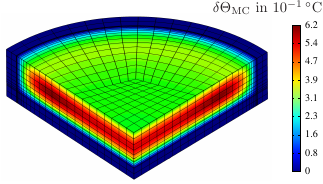}
	\label{fig:mc_Th_3}
	}
	\\
	\subfloat
	[$t = \SI{46800}{\s}$]
	{
	\includegraphics[width=0.4\textwidth]{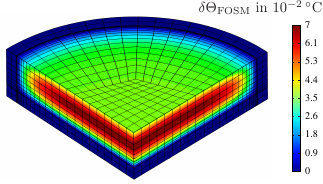}
	\label{fig:fosm_Th_5}
	}
	\hspace{0.04\textwidth}
	\subfloat
	[$t = \SI{46800}{\s}$]
	{
	\includegraphics[width=0.4\textwidth]{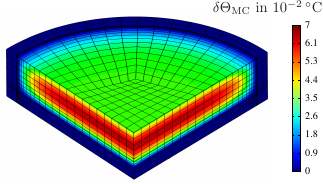}
	\label{fig:mc_Th_5}
	}
	\caption{Uncertainty in nodal temperatures during curing quantified with first-order second-moment method (left) and Monte Carlo method (right)}
	\label{fig:uq_Theta_cylinder}
\end{figure}

\begin{figure}[h!]
	\centering
	\subfloat
	[$t = \SI{29040}{\s}$]
	{
	\includegraphics[width=0.4\textwidth]{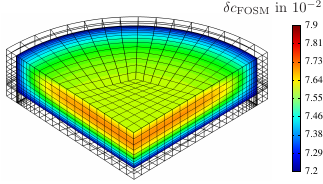}
	\label{fig:fosm_c_2}
	}
	\hspace{0.04\textwidth}
	\subfloat
	[$t = \SI{29040}{\s}$]
	{
	\includegraphics[width=0.4\textwidth]{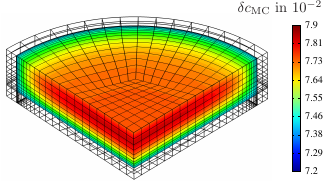}
	\label{fig:mc_c_2}
	}
	\\
	\subfloat
	[$t = \SI{29400}{\s}$]
	{
	\includegraphics[width=0.4\textwidth]{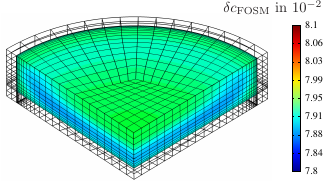}
	\label{fig:fosm_c_3}
	}
	\hspace{0.04\textwidth}
	\subfloat
	[$t = \SI{29400}{\s}$]
	{
	\includegraphics[width=0.4\textwidth]{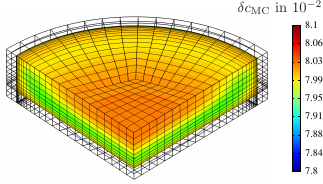}
	\label{fig:mc_c_3}
	}
	\\
	\subfloat
	[$t = \SI{46800}{\s}$]
	{
	\includegraphics[width=0.4\textwidth]{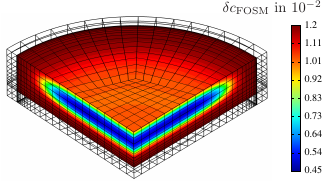}
	\label{fig:fosm_c_5}
	}
	\hspace{0.04\textwidth}
	\subfloat
	[$t = \SI{46800}{\s}$]
	{
	\includegraphics[width=0.4\textwidth]{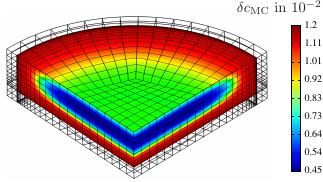}
	\label{fig:mc_c_5}
	}
	\caption{Uncertainty in degree of cure quantified with first-order second-moment method (left) and Monte Carlo method (right)}
	\label{fig:uq_c_cylinder}
\end{figure}

Below, the results for the three uncertainty quantification studies described above are provided and discussed.
\paragraph{Case I - Uncertain material parameters.} The nodal temperature uncertainty $\delta\mathrm{\Theta}$ estimated with FOSM (Eq.~\eqref{eq:varFOSM_model}) and Monte Carlo method (Eq.~\eqref{eq:varMC_model}) is visualized in Fig.~\ref{fig:uq_Theta_cylinder} for three particular time points (gray circles in Fig.~\ref{fig:curingPath_timeAdap}).
Additionally, the uncertainty $\delta c$ of the curing degree in the epoxy resin is illustrated in Fig.~\ref{fig:uq_c_cylinder}. 
Several notable aspects emerge from the quantified uncertainties of the two quantities of interest. First, the temperature uncertainty $\delta\mathrm{\Theta}$, shown in Fig.~\ref{fig:uq_Theta_cylinder}, is relatively small compared to the simulation results illustrated in Fig.~\ref{fig:sim_cylinder}, as indicated by its order of magnitude. In contrast, the degree of cure exhibits larger uncertainties, as evident when comparing Figs.~\ref{fig:fosm_c_2} and \ref{fig:fosm_c_3} with the corresponding simulation results in Figs.~\ref{fig:sim_c_2} and \ref{fig:sim_c_3}. Furthermore, the temporal evolution of the degree of cure uncertainty in Fig.~\ref{fig:uq_c_cylinder} reveals that the uncertainty level decreases towards the end of the curing process. Both findings are consistent with physical expectation: higher uncertainty is observed during curing, i.e., where rapid changes of the material state occur, while lower uncertainty is found once the curing process is almost finished (compare Fig.~\ref{fig:sim_c_5}).

Additionally, the temperature uncertainties quantified using FOSM and the Monte Carlo method show good agreement in Fig.~\ref{fig:uq_Theta_cylinder}. This demonstrates the applicability of FOSM, even for complex constitutive relations in a transient setting involving nonlinearities in both physics (due to radiation) and constitutive behavior. However, it is important to remember that FOSM is a first-order approximation, which explains the minor deviations observable in Fig.~\ref{fig:uq_Theta_cylinder}. In contrast, the degree of cure uncertainties shown in Fig.~\ref{fig:uq_c_cylinder} initially appear to exhibit larger differences between the two methods. However, this visual impression can be misleading, as the range between the minimum and maximum values is comparably small. Taking this into consideration, the difference in the degree of cure uncertainty is typically on the order of $10^{-3}$. This level of discrepancy is to be expected when using FOSM, given its first-order approximation and the fact that nonlinear effects and non-Gaussian distributions for some input parameters are only partially resolved.

\begin{figure}[ht]
	\centering
	\subfloat
	[Temperature]
	{
	\includegraphics[width=0.7\textwidth]{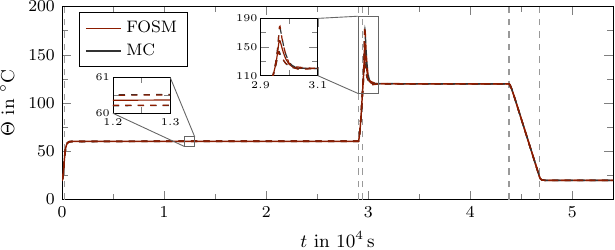}
	\label{fig:modelRespUncert_theta}
	}
	\\
	\subfloat
	[Degree of cure]
	{
	\includegraphics[width=0.7\textwidth]{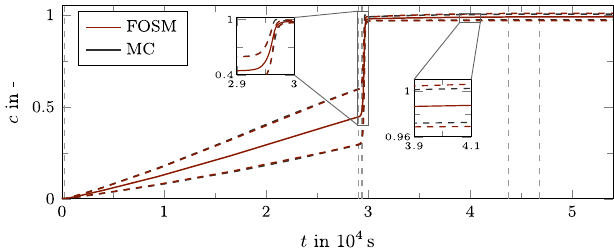}
	\label{fig:modelRespUncert_c}
	}
	\caption{Temporal evolution of model response (solid line) and uncertainty (dashed lines) in temperature and degree of cure due to uncertain material parameters. Uncertainties are visualized with confidence level $\SI{95}{\percent}$. Results are evaluated at the center of the top surface. Vertical lines indicate intervals of the curing path.}
	\label{fig:temporalEvol_ModelRespUncert}
\end{figure}

Since a quantitative comparison of spatial fields is difficult, especially regarding their temporal evolution, we evaluate the results of FOSM and the Monte Carlo method in greater detail at a single point on the top surface of the epoxy resin (indicated by a circle in Fig.~\ref{fig:meshBc}). The associated model response and uncertainties for both temperature $\mathrm{\Theta}$ and degree of cure $c$ are visualized in Fig.~\ref{fig:temporalEvol_ModelRespUncert}.
Apparently, the nodal temperature in Fig.~\ref{fig:modelRespUncert_theta} closely follows the profile of the prescribed temperature $\hat{\mathrm{\Theta}}$ (see Fig.~\ref{fig:curingPath_timeAdap}). However, a pronounced peak of the temperature is observed shortly after the onset of the post-curing phase at $t \approx \num{3.e04}$, which can be attributed to the exothermic curing reaction. This temperature increase coincides with a significant increase in the degree of cure, shown in Fig.~\ref{fig:modelRespUncert_c}. The aforementioned small temperature uncertainty $\delta\mathrm{\Theta}$ is also evident in Fig.~\ref{fig:modelRespUncert_theta} and results from the specific problem setup investigated here, since highly transient temperature effects occur only during the transition between pre- and post-curing phases. Nevertheless, Fig.~\ref{fig:modelRespUncert_theta} reveals that the FOSM can accurately capture the elevated level of uncertainty during periods of temperature increase.

In contrast, the uncertainty in the degree of cure is more pronounced, as visualized in Fig.~\ref{fig:modelRespUncert_c}. During the pre-curing phase, the degree of cure increases linearly, but rises rapidly once the post-curing phase at elevated temperature begins at $t = \SI{29040}{\s}$. The curing reaction is almost complete soon after the start of the post-curing phase. The results of the FOSM closely agree with those obtained from the Monte Carlo method during the pre-curing phase and the subsequent rapid increase in the degree of cure, with only minor deviations observed after the curing is completed. Compared to the model response, the importance of accounting for uncertainties is justified by the comparatively wide confidence interval during the pre-curing stage, clearly indicating the influence of material parameter uncertainties at this stage.

Furthermore, the temporal evaluation highlights the close agreement between the two methods, whereas the spatial visualization in Fig.~\ref{fig:uq_c_cylinder} may not suggest this at first sight. Overall, these results support the applicability of the FOSM for nonlinear problems involving considerable material parameter uncertainties (compare Tab.~\ref{tab:uncert}) and even those with non-Gaussian distributions. In terms of efficiency, we used $n_\mathrm{MC} = 300$ samples, i.e., full model evaluations, for the Monte Carlo method. In contrast, the FOSM with central difference quotients requires $35$ model evaluations ($1$ for the given parameters and $34$ for the difference quotients). The efficiency of the FOSM could be further improved by drawing on forward difference quotients or analytical derivatives of the model response, as demonstrated in \cite{troegerkulozikhartmann2025}. It should be emphasized that our focus is on assessing the applicability of FOSM for nonlinear constitutive relations, rather than maximizing computational efficiency. Nevertheless, the efficiency of FOSM comes at the cost of potential approximation errors.

\begin{figure}[ht]
	\centering
	\subfloat
	[Temperature]
	{
	\includegraphics[width=0.7\textwidth]{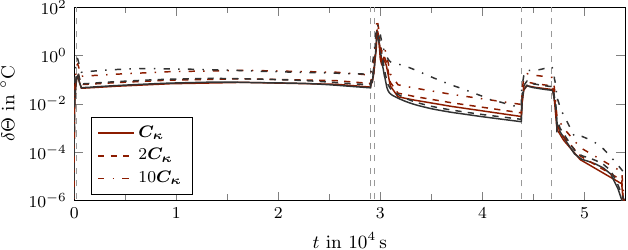}
	\label{fig:uq_theta_matpar}
	}
	\\
	\subfloat
	[Degree of cure]
	{
	\includegraphics[width=0.7\textwidth]{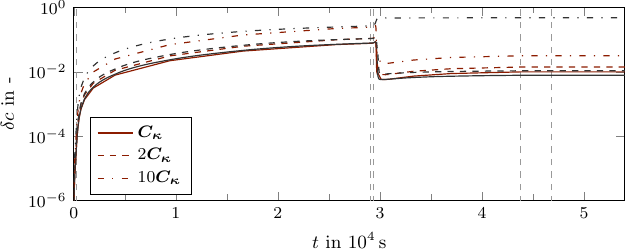}
	\label{fig:uq_c_matpar}
	}
	\caption{Temporal evolution of uncertainty in temperature and degree of cure due to uncertain material parameters with different levels of material parameter variance $(\delta\bm{\kappa}_\mathrm{MC})^2$. Comparison of FOSM (red) and the Monte Carlo method (black). Results are evaluated at the center of the top surface. Vertical lines indicate intervals of the curing path.}
	\label{fig:uq_theta_c_matpar}
\end{figure}

\paragraph{Case II - Scaled material parameter variance.} The second study examines how the difference between the results obtained with FOSM and those from the Monte Carlo method changes as the variance of the uncertain material parameters increases. As explained above, we simultaneously scale all parameter variances by a factor $k$ by inflating their empirical distributions. Since the model response at the mean parameter values is unaffected by variance scaling, we report only the uncertainties $\delta\mathrm{\Theta}$ and $\delta c$. These are presented in Fig.~\ref{fig:uq_theta_c_matpar} for both a twofold and tenfold increase in material parameter variance, alongside the original results from the previous analysis.
Both uncertainties exhibit increased uncertainty levels as the material parameter variances are scaled up. Nevertheless, the results from FOSM and the Monte Carlo method remain in close agreement throughout most of the curing path. A notable exception arises at the beginning of the post-curing phase, where the temporal course of the temperature uncertainty $\delta\mathrm{\Theta}$ changes with increasing variance, as shown in Fig.~\ref{fig:uq_theta_matpar}. This change is not captured by the FOSM, since the method keeps the sensitivities fixed and only scales the parameter variances.

Another notable exception is the degree of cure under a tenfold variance scaling. After the onset of the post-curing phase at $t = \SI{2.904e04}{\s}$, the uncertainty in the degree of cure $\delta c$ estimated using the Monte Carlo method remains high, at approximately $\num[round-precision=2]{0.45}$. Although this effect is not captured by the FOSM, it is important to recall that the degree of cure attains values in $[0,1]$. The source of this large uncertainty in the Monte Carlo results with a tenfold increase in variance can be attributed to the fact that the parameter variations are so substantial that, for some parameter samples, no curing occurs at all. In contrast, other parameter samples lead to full curing with the degree of cure approaching one. Conclusively, the FOSM remains applicable even for larger material parameter variances, as long as the results continue to be physically meaningful for the material under investigation.

\begin{figure}[ht]
	\centering
	\subfloat
	[FOSM]
	{
	\includegraphics[width=0.4\textwidth]{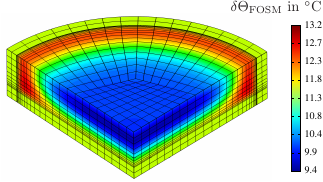}
	\label{fig:fosm_Th_bc_3}
	}
	\hspace{0.04\textwidth}
	\subfloat
	[Monte Carlo method]
	{
	\includegraphics[width=0.4\textwidth]{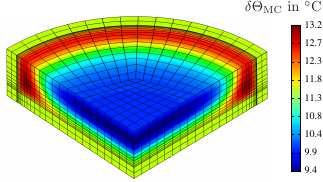}
	\label{fig:mc_Th_bc_3}
	}
	\\
	\subfloat
	[FOSM]
	{
	\includegraphics[width=0.4\textwidth]{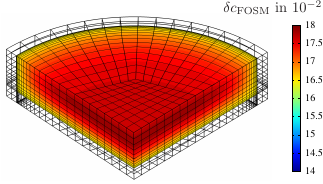}
	\label{fig:fosm_c_bc_3}
	}
	\hspace{0.04\textwidth}
	\subfloat
	[Monte Carlo method]
	{
	\includegraphics[width=0.4\textwidth]{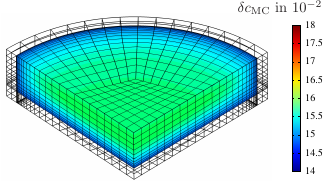}
	\label{fig:mc_c_bc_3}
	}
	\caption{Uncertainties of nodal temperatures (top) and degree of cure (bottom) at $t = \SI{29400}{\s}$ due to uncertain Dirichlet and mixed boundary conditions.}
	\label{fig:uncert_bc_fullSim}
\end{figure}

\paragraph{Case III - Uncertain boundary conditions.} In contrast to the previous two case studies, this third study focuses on uncertain boundary conditions while treating the material parameters as deterministic. Two aspects are assessed: first, the model response uncertainty resulting from uncertain Dirichlet and mixed boundary conditions; second, the influence of uncertain mixed boundary conditions only. Uncertain Dirichlet boundary conditions are introduced by varying seven characteristic temperature values $\hat{\mathrm{\Theta}}_j$, $j=1,\ldots,7$, on the curing path, as explained earlier. Uncertainty in the mixed boundary conditions is considered by treating the convection coefficient $h$ and the surface emissivity $\varepsilon$ as random variables. The spatial distributions of both temperature uncertainty and degree of cure uncertainty for the case of uncertain Dirichlet and mixed boundary conditions are visualized in Fig.~\ref{fig:uncert_bc_fullSim}.
\sisetup{round-precision=0}
The time point $t=\SI{29400}{\s}$ is depicted in Fig.~\ref{fig:uncert_bc_fullSim}, as it corresponds to the state with the highest uncertainty. The temperature uncertainty $\delta\mathrm{\Theta}$ is significantly larger (on the order of $10^1$) compared to the case with uncertain material parameters, which was on the order of $10^{-1}$, see Fig.~\ref{fig:uq_Theta_cylinder}. This substantial increase can be clearly attributed to the uncertain Dirichlet boundary conditions, as evidenced by the non-zero temperature uncertainty of the aluminum container. In contrast, the aluminum container exhibited no temperature uncertainty in previous analyses with deterministic Dirichlet boundary conditions. Nevertheless, the results of FOSM and the Monte Carlo method still agree very well for the temperature field.

\begin{figure}[ht]
	\centering
	\subfloat
	[Temperature]
	{
	\includegraphics[width=0.7\textwidth]{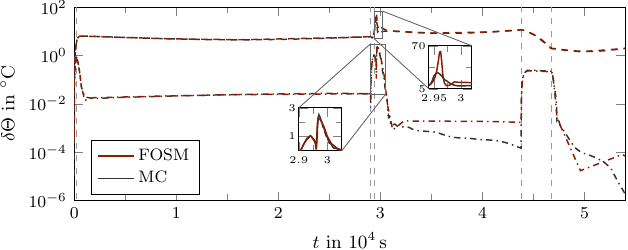}
	\label{fig:uq_theta_bc}
	}
	\\
	\subfloat
	[Degree of cure]
	{
	\includegraphics[width=0.7\textwidth]{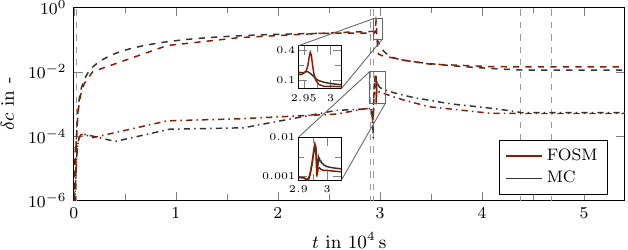}
	\label{fig:uq_c_bc}
	}
	\caption{Temporal evolution of uncertainty in temperature and degree of cure due to uncertain boundary conditions. Uncertainties are illustrated for uncertain Dirichlet and mixed boundary conditions (dashed lines) and only uncertain mixed boundary conditions (dash-dotted lines). Results are evaluated at the center of the top surface. Vertical lines indicate intervals of the curing path.}
	\label{fig:uq_theta_c_bc}
\end{figure}

The spatial field of the degree of cure uncertainty $\delta c$ is shown in Figs.~\ref{fig:fosm_c_bc_3} and \ref{fig:mc_c_bc_3}. In contrast to the temperature uncertainty, FOSM yields a higher uncertainty estimate for the degree of cure than the Monte Carlo method. This difference is spatially uniform with a magnitude of approximately $\num[round-precision=0]{2.e-03}$ and can be related to the first-order approximation of the FOSM. The temporal evolution of both uncertainties at a single point on the top surface is visualized in Fig.~\ref{fig:uq_theta_c_bc}.
Overall, the FOSM again shows close agreement with the Monte Carlo method for both uncertainties $\delta\mathrm{\Theta}$ and $\delta c$. However, the onset of the post-curing phase, specifically $t = \num[round-precision=1]{2.9}\ldots\SI[round-precision=0]{3.e04}{\s}$, is not captured accurately when both Dirichlet and mixed boundary conditions are uncertain. This limitation is evident for both temperature and degree of cure uncertainties, as shown in the magnifications in Fig.~\ref{fig:uq_theta_c_bc}. At this stage, the first-order approximation provided by FOSM appears to be insufficient, resulting in the observed discrepancy. Nevertheless, the overall temporal course of the uncertainties is very well represented by FOSM, further demonstrating the applicability even in the presence of uncertain boundary conditions.

As a final evaluation, the impact of having only uncertain mixed boundary conditions is assessed. The resulting uncertainties are shown as dash-dotted lines in Fig.~\ref{fig:uq_theta_c_bc}. Once again, FOSM and the Monte Carlo method are in very close agreement, aside from minor deviations at later stages of the curing path. Remarkably, the overall level of uncertainty is up to three orders of magnitude lower compared to the scenario where both Dirichlet and mixed boundary conditions are uncertain. This highlights the significant influence of uncertain Dirichlet boundary conditions on the model response during numerical simulations, confirming similar findings in the context of mechanical analyses \cite{troegersartortigarhuomduesterhartmann2024}. Furthermore, these results emphasize the importance of accurately prescribing Dirichlet boundary conditions even during experiments, as such uncertainties can greatly affect the material behavior and hinder meaningful comparison between experiments and simulations. In contrast, the coefficients of mixed boundary conditions exert only a minor influence and could be justifiably treated as deterministic parameters in similar contexts.

\section{Summary and Conclusions}
\label{sec:conclusions}
In this study, we investigate uncertainty quantification in both calibration and simulation of the thermo-chemical curing of epoxy resins. The thermo-chemical curing behavior is modeled using several highly nonlinear constitutive relations. Calibration is conducted with the nonlinear least-squares method, whereas material parameter uncertainties are quantified based on the asymptotic properties of the nonlinear least-squares estimator. During the multi-step calibration procedure, uncertainty propagation is addressed using the first-order second-moment method, where a notable increase in uncertainty occurs due to propagation effects. A comparison of the results with material parameters and associated uncertainties obtained from the Monte Carlo method demonstrates reasonable agreement for the material parameters. Larger deviations arise for parameters exhibiting clearly non-Gaussian distributions, which the first-order second-moment method can not adequately resolve. The robustness of the uncertainty propagation using the first-order second-moment method is validated with several coverage tests, highlighting that the detrimental effect of calibration steps based on sparse data should not be underestimated in multi-step model calibration schemes.

After quantifying the material parameter uncertainties, we assess their influence on the model response, specifically, nodal temperatures and degree of cure, using both the first-order second-moment method and the Monte Carlo method. The results indicate that the uncertainty quantification yields valuable insights and that both methods display good agreement. As expected, not all nonlinear effects are fully captured by the first-order second-moment method due to its first-order approximation. Consequently, the first-order second-moment method is very efficient for uncertainty quantification and provides reasonable results even in a nonlinear transient setting involving highly nonlinear constitutive relations with multiple uncertain parameters, whereas potential approximation errors must be kept in mind. Importantly, the flexibility of the first-order second-moment method extends beyond the propagation of material parameter uncertainties, allowing its application to uncertain boundary conditions as well. Our results show that uncertainties in the Dirichlet boundary conditions can severely increase the model response uncertainty. Therefore, efforts to reduce such uncertainties should be given priority.

\paragraph{Acknowledgements.}
This work was funded by the Deutsche Forschungsgemeinschaft (DFG, German Research Foundation) -- grant DFG HA 2024/20-1 (project 516997390). Furthermore, the authors gratefully acknowledge Dr.-Ing. Chris Leistner for granting access to the experimental data.

\bibliographystyle{ieeetr}
\bibliography{literature}

\end{document}